\pdfoutput=1
\documentclass[11pt,a4paper]{article}
\usepackage{jheppub,color,amsfonts,mathrsfs}
\usepackage[caption=false]{subfig}

\newcommand{\beq}{\begin{eqnarray}}
\newcommand{\eeq}{\end{eqnarray}}
\newcommand{\non}{\nonumber\\}

\newcommand{\p}{\partial}
\DeclareMathOperator{\U}{U}
\DeclareMathOperator{\SU}{SU}

\DeclareMathOperator{\SO}{SO}

\DeclareMathOperator{\diag}{diag}

\DeclareMathOperator{\Tr}{Tr}

\DeclareMathOperator{\sign}{sign}

\DeclareMathOperator{\eom}{eom}

\newcommand{\twonorm}[1]{|\mkern-1mu|#1|\mkern-1mu|}

\renewcommand{\i}{\mathrm{i}}
\renewcommand{\d}{{\mathrm{d}}}
\renewcommand{\L}{{\rm L}}
\newcommand{\R}{{\rm R}}
\newcommand{\CP}{\mathbb{C}P}
\newcommand{\sfe}{{\sf e}}
\newcommand{\calE}{\mathcal{E}}

\newcommand{\figsfolder}{figs/}

\newtheorem{theorem}{Theorem}

\newtheorem{lemma}[theorem]{Lemma}

\title{Chiral non-Abelian vortex molecules in dense QCD}

\author{Sven Bjarke Gudnason$^1$,}
\affiliation{$^1$Institute of Contemporary Mathematics, School of
  Mathematics and Statistics, Henan University, Kaifeng, Henan 475004,
  P.~R.~China}
\emailAdd{gudnason(at)henu.edu.cn}
\author{Muneto Nitta$^{2,3}$}
\affiliation{$^2$Department of Physics \& Research and Education Center for Natural Sciences,
Keio University, Hiyoshi 4-1-1, Yokohama, Kanagawa 223-8521, Japan}
\affiliation{$^3$International Institute for Sustainability with Knotted Chiral Meta Matter (SKCM$^2$),
Hiroshima University, 1-3-2 Kagamiyama, Higashi-Hiroshima, Hiroshima 739-8511, Japan}
\emailAdd{nitta(at)phys-h.keio.ac.jp}

\abstract{
The color-flavor locked phase of high density QCD admits non-Abelian vortices, that are 
 superfluid vortices with confined color-magnetic fluxes.
Vortex solutions were thus far constructed in an axially symmetric Ansatz 
with common vortex winding in 
the left- and 
right-handed diquark condensates.
 In this paper, 
we explore vortex solutions without any Ansatz.
In addition to axially symmetric configurations known before, we find that 
the axial symmetry is broken 
in certain parameter regions;
in one case a single vortex
is split into two chiral non-Abelian vortices, 
i.e.~vortices with 
winding only in 
the left or
right-handed diquark condensate, 
and they are connected by one or two domain walls 
forming a non-Abelian vortex molecule.
In the other case, 
 a chiral non-Abelian vortex molecule is confined inside a domain wall elongated to spatial infinity. 
}

\begin{document}
\maketitle

\section{Introduction}
Exploring states of matter 
in extreme conditions is 
an important problem 
in modern physics. 
In particular, 
nuclear and quark matter 
in extreme conditions 
such as high density, strong magnetic field, and rapid rotation 
are considered to be 
pivotal for a full understanding of the interior of compact stars such as neutron stars. 
The quark matter described by quantum chromodynamics (QCD) at
high densities and low temperatures 
is expected to exhibit color superconductivity as well as superfluidity~\cite{Alford:2007xm}. 
In the limit of extremely high density, a phase with  
 three-flavor symmetric matter called 
 the color-flavor locked (CFL) phase 
is realized~\cite{Alford:1998mk}, 
while 
the two-flavor superconducting (2SC) phase~\cite{Alford:1997zt, Rapp:1997zu}
is also proposed 
for the intermediate-density region of the QCD phase diagram.
In metallic superconductors and superfluids, quantum vortices or magnetic flux tubes play significant roles in color-superconducting quark matter, 
as reviewed in ref.~\cite{Eto:2013hoa}. 
In the CFL phase, 
the most fundamental stable vortices are 
non-Abelian vortices carrying color-magnetic fluxes and
1/3 superfluid circulation~\cite{Eto:2013hoa,Balachandran:2005ev, Nakano:2007dr, Nakano:2008dc,
  Eto:2009kg, Eto:2009bh, Eto:2009tr}.\footnote{
  Similar non-Abelian vortices were studied in 
  supersymmetric QCD \cite{Hanany:2003hp,Auzzi:2003fs,
Hanany:2004ea,Shifman:2004dr,Eto:2004rz,Eto:2005yh} 
(see refs.~\cite{Tong:2005un,Eto:2006pg,Shifman:2007ce,Shifman:2009zz} for a review) 
and in two-Higgs doublet models 
(as an extension of the Standard Model) \cite{Eto:2018hhg,Eto:2018tnk,
Eto:2019hhf,Eto:2020hjb,Eto:2020opf,Eto:2021dca}.
 They are different from the system studied here, in the sense that 
 the overall $\U(1)$ symmetry is gauged in supersymmetric QCD, 
 and it is an approximate 
 symmetry in two-Higgs doublet models.
}
Superfluid vortices 
without fluxes called Abelian vortices
\cite{Forbes:2001gj,Iida:2002ev}  
are dynamically unstable to decay into 
three non-Abelian vortices  
\cite{Nakano:2007dr,Cipriani:2012hr,Alford:2016dco}. 
  A non-Abelian vortex confines bosonic and fermionic gapless modes in its core; 
  the former are Nambu-Goldstone 
   ${\mathbb C}P^2 (\simeq {\rm SU}(3)/[{\rm SU}(2)\times {\rm U}(1)])$ modes originated from 
  spontaneous breaking of the CFL symmetry in the vicinity of the core
\cite{Nakano:2007dr,Eto:2009bh,Eto:2013hoa,Eto:2009tr} while the latter are
  three Majorana fermions  \cite{Yasui:2010yw,Fujiwara:2011za,Chatterjee:2016ykq}. 
  Under a rapid rotation, there appear a huge number 
  of vortices 
  (about $10^{19}$ for typical neutron stars) aligned along the rotation axis, 
  forming a vortex lattice with the lattice spacing of the order of a micrometer \cite{Kobayashi:2013axa,Hirono:2012ki}.
Such vortices penetrate via crossover between the CFL phase and hyperon nuclear matter 
within a quark-hadron continuity~\cite{Cipriani:2012hr,Alford:2018mqj, Chatterjee:2018nxe,
  Chatterjee:2019tbz,  Cherman:2018jir, Hirono:2018fjr, Hirono:2019oup, Cherman:2020hbe}
  without \cite{Alford:2018mqj} or with 
  Boojums \cite{Cipriani:2012hr,Chatterjee:2018nxe,Chatterjee:2019tbz}. 
  This direction has stimulated further studies of the Higgs-confinement continuity in the presence of a vortex \cite{Cherman:2020hbe,Hayashi:2023sas,Cherman:2024exo,Hayata:2024nrl}.

The CFL phase is characterized by the two diquark condensates 
of left and right-handed quarks $q_{\rm L,R}$,
$(\Phi_{\rm L,R})_{\alpha a} \sim \epsilon_{\alpha \beta \gamma} 
\epsilon_{abc} q_{\rm L,R}^{\beta b}  q_{\rm L,R}^{\gamma c}$ 
with the color indices $\alpha,\beta,\gamma=r,g,b$ 
and the flavor indices $a,b,c=u,d,s$.   
In the ground state, they
simultaneously develop vacuum expectation values (VEVs)
as $\Phi_{\rm L} = - \Phi_{\rm R}$. 
In the previous studies \cite{Balachandran:2005ev,Eto:2009kg,Eto:2013hoa}, 
non-Abelian vortices were constructed 
in a simplified setup with the assumption
 $\Phi_{\rm L} = - \Phi_{\rm R}$ 
 in the entire region, including vortex cores 
 and  
the axisymmetry around the vortices. 
In order to investigate more general vortex configurations, 
it is convenient to
recall a simpler setup of condensed matter systems with 
simultaneous condensation of two components
$\Phi_1$ and $\Phi_2$. 
Examples contain two-gap superconductors 
 \cite{Babaev:2001hv,
 doi:10.1143/JPSJ.70.2844,
PhysRevLett.88.017002,Goryo_2007},
chiral $P$-wave superconductors \cite{PhysRevLett.80.5184,PhysRevB.86.060514} 
and 
two-component Bose-Einstein condensates (BECs)
\cite{
Son:2001td,
Kasamatsu:2004tvg,
Cipriani:2013nya,
Tylutki:2016mgy,
Eto:2017rfr,
Eto:2019uhe,Kobayashi:2018ezm}, 
in which the overall phase is a gauge (global) symmetry 
in the superconductors (BECs), 
and the relative phase is an approximate global symmetry explicitly broken by 
a Josephson term (Rabi coupling) 
$\Phi_1^* \Phi_2 + {\rm c.c.}$ 
(or $\Phi_1^{*2} \Phi_2^2+ {\rm c.c.}$ for chiral $P$-wave superconductors \cite{PhysRevLett.80.5184,PhysRevB.86.060514}).
A singly-quantized vortex 
has winding in both components: $\Phi_1= \Phi_2 \sim e^{\i\theta}$ at the same position 
with the azimuthal angle $\theta$. Depending on the situation, 
it 
can be split into 
two half-quantized vortices 
 $(\Phi_1,\Phi_2 )\sim (e^{\i\theta},1)$ 
and $(\Phi_1,\Phi_2 )\sim (1, e^{\i\theta})$ at different positions connected by 
a sine-Gordon soliton 
 \cite{doi:10.1143/JPSJ.70.2844,
PhysRevLett.88.017002,Goryo_2007,Son:2001td} 
(or two domain walls for chiral $P$-wave superconductors \cite{PhysRevB.86.060514}) 
forming a vortex molecule.\footnote{
See ref.~\cite{Chatterjee:2019jez} for the case of more general higher-order Josephson-like terms 
and 
refs.~\cite{Eto:2012rc,Eto:2013spa} for the cases of more components 
and corresponding 
vortex-molecule structures. 
}

In this paper, 
we explore non-Abelian vortex solutions 
without imposing an Ansatz; 
no assumption of $\Phi_{\rm L} = - \Phi_{\rm R}$ 
nor of axisymmetry,
unlike the previous studies~\cite{Balachandran:2005ev, Eto:2009kg, Eto:2013hoa}. 
We find that 
for certain parameter choices,
a single non-Abelian vortex is split into 
two vortex cores breaking the axisymmetry, 
each of which has a winding 
only in left $\Phi_{\rm L}$ or in the right $\Phi_{\rm R}$ condensates.
Such constituents are called 
chiral non-Abelian vortices 
because a condensate of only left or right chirality has a vortex-winding 
\cite{Eto:2021nle}.
Each chiral non-Abelian vortex is attached by 
one or two domain walls \cite{Eto:2013hoa,Eto:2013bxa}, 
due to the fact that 
axial and chiral symmetries are explicitly 
broken by the mass and axial-anomaly terms. The two chiral non-Abelian vortices with opposite chiralities  
are connected by one or two domain walls 
forming a non-Abelian vortex molecule.
In certain parameter regions, the energy remains axisymmetric, but the vortex centers are not coincident -- the axial symmetry is hence broken.
We also find, in another parameter region, that the two chiral non-Abelian vortices are confined on a single domain wall, with 
one domain wall connecting them 
and two domain walls elongated to spatial infinities.

This paper is organized as follows.
In sec.~\ref{sec:CFL} we summarize 
the Ginzburg-Landau (GL) energy functional of the CFL phase of dense QCD 
and the ground states of the CFL phase.
In sec.~\ref{sec:vortices} we investigate vortex solutions without imposing an Ansatz, 
and find a non-Abelian vortex molecule 
in which two chiral vortices 
of opposite chiralities 
are connected by one or two domain walls, as well as a vortex molecule confined inside a single domain wall. 
Sec.~\ref{sec:summary} is devoted to a summary and discussion.

\section{Color-flavor-locked phase of 3-flavor dense QCD}\label{sec:CFL}

In this section, we review the GL 
model for the CFL phase 
to fix our notations, 
and summarize the ground state structures.

\subsection{Ginzburg-Landau model for the color-flavor locked phase}

Let $q_{\L,\R}$ be left- and right-handed quarks.
Then, the CFL phase is characterized by the simultaneous diquark condensates
of the left and right chiralities:
\beq
(\Phi_{\L,\R})_{\alpha a}\sim\epsilon_{\alpha\beta\gamma}\epsilon_{abc}q_{\L,\R}^{\beta b}q_{\L,\R}^{\gamma c},
\eeq
which are 3-by-3 matrices of scalar fields with $\alpha,\beta,\gamma=r,g,b=1,2,3$ being color indices and $a,b,c=u,d,s=1,2,3$ being flavor indices. 
The 
${\rm SU}(3)_{\rm C}$ color symmetry and 
${\rm U}(1)_{\rm B}$ baryon symmetry 
are exact while 
the 
${\rm SU}(3)_\L \times 
{\rm SU}(3)_\R$ chiral symmetry 
and ${\rm U}(1)_{\rm A}$ axial symmetry 
are approximate.  These symmetries 
act on the condensates as
\beq
 && \Phi_{\rm L} \to 
 e^{i\theta_{\rm B} +i\theta_{\rm A}} g_{\rm C} \Phi_{\rm L} U_{\rm L}^\dagger , \quad
  \Phi_{\rm R} \to 
   e^{i\theta_{\rm B} - i \theta_{\rm A}} 
  g_{\rm C} \Phi_{\rm R} U_{\rm R}^\dagger \nonumber\\
  && g_{\rm C} \in {\rm SU}(3)_{\rm C}, \quad 
  U_{\rm L,R} \in {\rm SU}(3)_{\rm L,R}, \quad
   e^{i\theta_{\rm B}}\in {\rm U}(1)_{\rm B} ,\quad
    e^{i\theta_{\rm A}} \in {\rm U}(1)_{\rm A}. \label{eq:G-on-Phi}
\eeq
The vector symmetry ${\rm SU}(3)_{\rm L+R}$ defined  by the condition $U_{\rm L} = U_{\rm R}$
is a subgroup of the chiral symmetry 
${\rm SU}(3)_{\rm L} \times {\rm SU}(3)_{\rm R}$, and
the rest of the generators 
parametrize Nambu-Goldstone bosons 
for the chiral symmetry breaking 
as the coset space
\beq
[{\rm SU}(3)_{\rm L} \times {\rm SU}(3)_{\rm R}] / {\rm SU}(3)_{\rm L+R}
\simeq {\rm SU}(3).\nonumber
\eeq

The static Hamiltonian (energy functional) of the GL model takes the form\footnote{The $\lambda_2$ and $\lambda_3$ terms are divided by $N=3$ for convenience, compared e.g.~with ref.~\cite{Eto:2021nle}.}
\begin{align}
E &= \frac{1}{2g^2}\twonorm{F}^2 
+ \twonorm{\d_A\Phi_L}^2
+ \twonorm{\d_A\Phi_R}^2
+ V,\label{eq:E}\\
V &= -\frac{m^2}{2}\int_M\star\Tr\big(\Phi_\L^\dag\Phi_\L+\Phi_\R^\dag\Phi_\R\big)
+ \frac{\lambda_1}{4}\int_M\star\Tr\left[\big(\Phi_\L^\dag\Phi_\L\big)^2+\big(\Phi_\R^\dag\Phi_\R\big)^2\right]\non
&\phantom{=\ }
+ \frac{\lambda_2}{12}\int_M\star\left(\left(\Tr\big[\Phi_\L^\dag\Phi_L\big]\right)^2 + \left(\Tr\big[\Phi_\R^\dag\Phi_\R\big]\right)^2\right)
+ \frac{\lambda_3}{6}\int_M\star\Tr\big[\Phi_\L^\dag\Phi_\L\big]\Tr\big[\Phi_\R^\dag\Phi_\R\big]\non
&\phantom{=\ }
+ \frac{\lambda_4}{2}\int_M\star\Tr\big[\Phi_\R^\dag\Phi_\L\Phi_\L^\dag\Phi_\R\big]
+ \gamma_1\int_M\star\Tr\big(\Phi_\L^\dag\Phi_\R + \Phi_\L\Phi_\R^\dag\big)\non
&\phantom{=\ }
+ \gamma_2\int_M\star\Tr\left[\big(\Phi_\L^\dag\Phi_\R\big)^2+\big(\Phi_\R^\dag\Phi_\L\big)^2\right]
+ \gamma_3\int_M\star\left(\det\big(\Phi_\L^\dag\Phi_\R\big)+\det\big(\Phi_\R^\dag\Phi_\L\big)\right),\label{eq:V}
\end{align}
where the inner product on $M=\mathbb{R}^2$ 
(the plane orthogonal to vortices) is defined as
\beq
\langle X,Y\rangle := \Tr\int_M X^\dag\wedge\star Y,
\eeq
where $X,Y$ are both $r$-forms as well as 3-by-3 complex matrices and $\star$ is the Hodge star mapping $r$-forms to ($2-r$)-forms with the property $\star\star=(-1)^r$ (in two dimensions). 
The norm squared is then defined as
\beq
\twonorm{X}^2 := \langle X,X\rangle,
\eeq
and the integral of the Hodge dual of a scalar is simply the normal
integral with the volume form: $\int_M\star Z=\int_M Z\,\d^2x$ with
$Z$ a 0-form.
The field-strength 2-form 
for the SU(3)$_{\rm C}$ color gauge field 
is defined as
\beq
F = \d A - A\wedge A = \frac12F_{ij}\d x^{ij},
\eeq
with the short-hand notation $\d x^{ij}:=\d x^i\wedge\d x^j$, $i,j=1,2$ and the gauge covariant derivative
\beq
\d_A\Phi_{\L,\R} := d\Phi_{\L,\R} - A \Phi_{\L,\R},
\eeq
where $A=A_i\d x^i$ is an anti-Hermitian 1-form, i.e.~$A^\dag=-A$ and is $\mathfrak{su}(3)$-valued.
This implies that it is also traceless.

For convenience, we provide a few expressions in component form
\begin{align}
\frac{1}{2g^2}\twonorm{F}^2&=-\frac{1}{4g^2}\int_M \Tr\left[F_{ij}F^{ij}\right]\d^2x\geq0,\\
\twonorm{\d_A\Phi_\L}^2&=\int_M \Tr\left[(\p_i\Phi_\L-A_i\Phi_\L)^\dag (\p^i\Phi_\L-A^i\Phi_\L)\right]\d^2x,
\end{align}
where spatial indices $i,j$ are lowered (raised) by the (inverse) metric $g_{ij}=\delta_{ij}$ ($g^{ij}=\delta^{ij}$).
Note the negative sign in front of $F_{ij}F^{ij}$ is due to the anti-Hermitian gauge field, i.e.~$F^\dag=-F$.

The GL parameters in the GL model in eq.~\eqref{eq:V} were microscopically 
calculated in the asymptotically 
high-density region of QCD 
\cite{Iida:2000ha,Iida:2001pg,Giannakis:2001wz}. 
In this paper, we leave those parameters as free GL parameters.

\subsection{First variation}

In order to obtain the equations of motion, we perform a first variation of the energy functional.
Let $(A_\tau,\Phi_{\L,\tau},\Phi_{\R,\tau})$ be smooth variations of the fields $(A,\Phi_\L,\Phi_\R)$ for all $\tau$.
Denote by $\alpha=\p_\tau A_\tau|_{\tau=0}$, $\beta_\L=\p_\tau\Phi_{\L,\tau}|_{\tau=0}$ and $\beta_\R=\p_\tau\Phi_{\R,\tau}|_{\tau=0}$.
We thus obtain
\begin{align}
\frac{\d}{\d\tau}\bigg|_{\tau=0}\!E
&=\langle\beta_\L,\eom_{\Phi_\L}\rangle_{L^2(M)}
+\langle\eom_{\Phi_\L},\beta_\L\rangle_{L^2(M)}
+\langle\beta_\R,\eom_{\Phi_\R}\rangle_{L^2(M)}
+\langle\eom_{\Phi_\R},\beta_\R\rangle_{L^2(M)}\non
&\phantom{=\ }
+\langle\alpha,\eom_A\rangle_{L^2(M)}
-\frac{1}{g^2}\Tr\int_M\d\left(\alpha\wedge\star F\right)\non
&\phantom{=\ }
+\Tr\int_M\d\left(\beta_\L^\dag\star\d_A\Phi_\L
+\star\d_A\Phi_\L^\dag\beta_\L
+\beta_\R^\dag\star\d_A\Phi_\R
+\star\d_A\Phi_\R^\dag\beta_\R\right),
\label{eq:first_variation}
\end{align}
with the equations of motion
\begin{align}
\eom_{\Phi_\L}&=\delta_A\d_A\Phi_\L
-\frac{m^2}{2}\Phi_\L
+\frac{\lambda_1}{2}\Phi_\L\Phi_\L^\dag\Phi_\L
+\frac{\lambda_2}{6}\Phi_\L\Tr(\Phi_\L^\dag\Phi_\L)
+\frac{\lambda_3}{6}\Phi_\L\Tr(\Phi_\R^\dag\Phi_\R)\non
&\phantom{=\ }
+\frac{\lambda_4}{2}\Phi_\R\Phi_\R^\dag\Phi_\L
+\gamma_1\Phi_\R
+2\gamma_2\Phi_\R\Phi_\L^\dag\Phi_\R
+\gamma_3\Xi_\L,\label{eq:eom_PhiL}\\
\eom_{\Phi_\R}&=\delta_A\d_A\Phi_\R
-\frac{m^2}{2}\Phi_\R
+\frac{\lambda_1}{2}\Phi_\R\Phi_\R^\dag\Phi_\R
+\frac{\lambda_2}{6}\Phi_\R\Tr(\Phi_\R^\dag\Phi_\R)
+\frac{\lambda_3}{6}\Phi_\R\Tr(\Phi_\L^\dag\Phi_\L)\non
&\phantom{=\ }
+\frac{\lambda_4}{2}\Phi_\L\Phi_\L^\dag\Phi_\R
+\gamma_1\Phi_\L
+2\gamma_2\Phi_\L\Phi_\R^\dag\Phi_\L
+\gamma_3\Xi_\R,\label{eq:eom_PhiR}\\
\eom_A&=\frac{1}{g^2}\left[\delta F+\star A\wedge\star F-\star(\star F\wedge A)\right]
-\d_A\Phi_\L\Phi_\L^\dag+\Phi_\L\d_A\Phi_\L^\dag
-\d_A\Phi_\R\Phi_\R^\dag+\Phi_\R\d_A\Phi_\R^\dag\non
&=\bigg(-\frac{1}{g^2}\p^iF_{ij}+\frac{1}{g^2}[A^i,F_{ij}]
-(\p_j\Phi_\L-A_j\Phi_L)\Phi_\L^\dag
+\Phi_\L(\p_j\Phi_\L^\dag+\Phi_\L^\dag A_j)\non
&\phantom{=\bigg(\ }
-(\p_j\Phi_\R-A_j\Phi_R)\Phi_\R^\dag
+\Phi_\R(\p_j\Phi_\R^\dag+\Phi_\R^\dag A_j)
\bigg)\d x^j,\label{eq:eom_A}
\end{align}
which are two 0-forms and a 1-form that vanish when the equations of motion are satisfied.
In eq.~\eqref{eq:first_variation}, the last two terms, being total derivatives, vanish by the assumption of the smooth variations $\alpha$ and $\beta$ having compact support.\footnote{From a physical point of view, the finite-energy condition requires $F$ and $\d_A\Phi_{\L,\R}$ to vanish at spatial infinity and hence the total derivatives vanish.}
$\delta$ is the coderivative and $\delta_A$ is the gauge covariant coderivative; writing out the $\delta_A\d_A$ we obtain
\beq
\delta_A\d_A\Phi_\L&=-\star\d_A\star\d_A\Phi_\L
=-(\p_i-A_i)(\p^i-A^i)\Phi_\L.
\eeq
Finally, the matrices $\Xi_{\L,\R}$ which are the variation of the determinant, can be written as
\begin{align}
(\Xi_\L)_{\alpha a}&=\frac12\epsilon_{abc}\epsilon_{def}(\Phi_\R)_{\alpha d}(\Phi_\L^\dag\Phi_R)_{be}(\Phi_\L^\dag\Phi_\R)_{cf},\\
(\Xi_\R)_{\alpha a}&=\frac12\epsilon_{abc}\epsilon_{def}(\Phi_\L)_{\alpha d}(\Phi_\R^\dag\Phi_L)_{be}(\Phi_\R^\dag\Phi_\L)_{cf}.
\end{align}
Having the equations of motion in hand, we can read off the perturbative masses from the linearized equations:
\beq
m_\Phi=\frac{m}{\sqrt{2}}, \qquad
m_A=g\sqrt{2(v_\L^2+v_\R^2)},
\eeq
with the two VEVs $v_\L=\langle\Phi_\L\rangle$ and $v_\R=\langle\Phi_\R\rangle$ being determined by the ground state equations, see below.

\subsection{Ground states}

Let us assume that the ground states are given by diagonal matrices, which thus do not break the $\SU(3)_{\rm C+L+R}$ symmetry. 
We will further use the $\U(1)_{\rm A,B}$ symmetries to set the phases of the VEVs to zero (or $\pi$) if possible.

Let us consider the ground state with only $m\neq0$, $\lambda_1>0$, $\lambda_2>0$.
The ground state solution compatible with this condition is unique
\beq
\Phi_\L=-\Phi_\R=v\mathbf{1}_3, \qquad
v = \frac{m}{\sqrt{\lambda_1+\lambda_2}},\qquad
\langle V\rangle = -\frac{3m^4}{2(\lambda_1+\lambda_2)},
\eeq
where $\langle V\rangle$ is the potential value at the VEV. 
The symmetry is broken to 
the CFL symmetry ${\rm SU}(3)_{\rm C+L+R}$
given by $g_{\rm C}=U_\L=U_\R$ in 
eq.~(\ref{eq:G-on-Phi}).

Turning on the mixed Hermitian terms, $\lambda_3>0$ and $\lambda_4>0$, two competing ground states appear:
\begin{align}
\Phi_\L=-\Phi_\R=u\mathbf{1}_3, \quad u=\frac{m}{\sqrt{\lambda_1+\lambda_2+\lambda_3+\lambda_4}},\quad
\langle V\rangle=-\frac{3m^4}{2(\lambda_1+\lambda_2+\lambda_3+\lambda_4)},
\end{align}
and
\beq
\left\{\begin{array}{l}
\Phi_\L=v\mathbf{1}_3\\\Phi_\R=\mathbf{0}_3
\end{array}\right\}
\ \ \textrm{or}\ \ 
\left\{\begin{array}{l}
\Phi_\L=\mathbf{0}_3\\\Phi_\R=v\mathbf{1}_3
\end{array}\right\},\quad 
v=\frac{m}{\sqrt{\lambda_1+\lambda_2}},\quad
\langle V\rangle=-\frac{3m^4}{4(\lambda_1+\lambda_2)}.
\eeq
The condition for the ground state being either the $v$ or the $u$ ground state is
\begin{align}
\lambda_1+\lambda_2 > \lambda_3+\lambda_4\ &: \qquad \Rightarrow \qquad
\Phi_\L=-\Phi_\R=u\mathbf{1}_3,\\
\lambda_1+\lambda_2 < \lambda_3+\lambda_4\ &: \qquad \Rightarrow \qquad
\left\{\begin{array}{l}
\Phi_\L=v\mathbf{1}_3\\\Phi_\R=\mathbf{0}_3
\end{array}\right\}
\quad\textrm{or}\quad
\left\{\begin{array}{l}
\Phi_\L=\mathbf{0}_3\\\Phi_\R=v\mathbf{1}_3
\end{array}\right\}.
\end{align}
The former is still the conventional CFL ground state with the CFL symmetry, while  
the latter ground state of the $v$-ground state (requiring large $\lambda_3$ or $\lambda_4$) contains two degenerate ground states and hence a domain wall that connects them.

We will now turn on $\gamma_1\neq0$.
There will still be two competing ground states that depend on whether $\lambda_3$ or $\lambda_4$ (or both) is large or not.
Due to the linear term (proportional to $\gamma_1$) in the ground state equations, the vanishing VEV from before is shifted slightly. 
For convenience we include $\gamma_2$ although it is allowed to vanish in the following ground state.
In particular, we get
\begin{align}
\Phi_\L&=-\sign(\gamma_1)\Phi_\R=w\mathbf{1}_3, \quad
w=\sqrt{\frac{m^2+2|\gamma_1|}{\lambda_1+\lambda_2+\lambda_3+\lambda_4+4\gamma_2}},\non
\langle V\rangle&=-\frac{3(m^2+2|\gamma_1|)^2}{2(\lambda_1+\lambda_2+\lambda_3+\lambda_4+4\gamma_2)},
\end{align}
when the condition
\beq
2\left(1+\frac{2|\gamma_1|}{m^2}\right)^2 > \left(1+\frac{\lambda_3+\lambda_4}{\lambda_1+\lambda_2}\right)\left(
1+\frac{2\gamma_1^2(\lambda_1+\lambda_2)}{m^4(\lambda_3+\lambda_4-\lambda_1-\lambda_2+4\gamma_2)}\right),
\label{eq:gamma12cond}
\eeq
is satisfied and otherwise the following is the ground state
\begin{align}
\Phi_\L&=w_\pm,\quad
\Phi_\R=-\sign(\gamma_1)w_\mp,\non
w_\pm&=\frac{m}{\sqrt{\lambda_1+\lambda_2}}\sqrt{\frac12\pm
\epsilon\sqrt{1-\frac{16\gamma_1^2(\lambda_1+\lambda_2)^2}{m^4(\lambda_3+\lambda_4-\lambda_1-\lambda_2+4\gamma_2)^2}}},\non
\langle V\rangle&=-\frac{3m^4}{4(\lambda_1+\lambda_2)}-\frac{6\gamma_1^2}{\lambda_3+\lambda_4-\lambda_1-\lambda_2+4\gamma_2},
\end{align}
with the sign $\epsilon=\sign[\lambda_3+\lambda_4-\lambda_1-\lambda_2+4\gamma_2]$.
The condition \eqref{eq:gamma12cond} choosing between the two types of ground states clearly gets complicated by the presence of $\gamma_1$ -- the Josephson term, but the ground state structure is changed only little by $\gamma_2$.
Interestingly, $\gamma_2$ does not affect the ground state condition when $\gamma_1=0$.
One can check that setting $\gamma_1:=0$, the condition \eqref{eq:gamma12cond} reduces to the previous condition, i.e.~$\lambda_1+\lambda_2>\lambda_3+\lambda_4$.

Clearly the ground state structure becomes only more complicated by turning on $\gamma_3\neq0$.
We will focus on the simplest case where we turn on $\gamma_3\neq0$ but leave $\gamma_1=\gamma_2=0$.
In this case, there is only a single ground state
\begin{align}
\Phi_\L&=-\Phi_\R=r\mathbf{1}_3, \qquad
r=\frac12\sqrt{\frac{\lambda_{1234}-\sqrt{\lambda_{1234}^2-8m^2|\gamma_3|}}{|\gamma_3|}}, \non
\langle V\rangle &=-\frac{\left(\lambda_{1234}-\sqrt{\lambda_{1234}^2-8m^2|\gamma_3|}\right)\left(16m^2|\gamma_3|-\lambda_{1234}\left(\lambda_{1234}-\sqrt{\lambda_{1234}^2-8m^2|\gamma_3|}\right)\right)}{32\gamma_3^2},
\end{align}
provided that $\gamma_3$ is small enough:
\beq
|\gamma_3|<\frac{\lambda_{1234}^2}{8m^2},
\eeq
and we have defined $\lambda_{1234}:=\sum_{i=1}^4\lambda_i$.
If on the other hand, the above condition for $|\gamma_3|$ is not satisfied, the ground state becomes that of the partially unbroken phase:
\beq
\left\{\begin{array}{l}
\Phi_\L=v\mathbf{1}_3\\\Phi_\R=\mathbf{0}_3
\end{array}\right\}
\ \ \textrm{or}\ \ 
\left\{\begin{array}{l}
\Phi_\L=\mathbf{0}_3\\\Phi_\R=v\mathbf{1}_3
\end{array}\right\},\quad 
v=\frac{m}{\sqrt{\lambda_1+\lambda_2}},\quad
\langle V\rangle=-\frac{3m^4}{4(\lambda_1+\lambda_2)}.
\eeq
A comment in store about the $\gamma_3$ term, is that the potential theoretically has runaway directions that can be triggered for very large field values. 
This must however physically be just an artifact of the low-energy EFT.

\section{Chiral vortex molecules}\label{sec:vortices}

In this section, we numerically construct 
non-axisymmetric vortex configurations 
such as chiral vortex molecules 
and chiral vortices confined on a domain wall.

\subsection{Vortex Ans\"atze for initial conditions}\label{sec:ansatze}

We seek solutions that describe vortex molecules that hence do not possess axial symmetry.
Nevertheless, the initial configurations for our simulation need an Ansatz for each of the two vortices, which will be detailed here.
The chiral $(1,0)$ vortex is given by
\begin{equation}
\Phi_\L=v_\L\diag\left(f(r)e^{\i\theta},1,1\right),\qquad
\Phi_\R=v_\R\diag(1,1,1),\qquad
A = -\i\epsilon_{ij}\frac{x^j}{2r^2}a(r)T \d x^i,\label{eq:10vortex_ansatz}
\end{equation}
with the matrix $T=\diag(\tfrac23,-\tfrac13,-\tfrac13)$ and the VEV of $\Phi_\L$ ($\Phi_\R$) being $v_\L$ ($v_\R$), depending on the ground state in question.
Similarly, the chiral $(0,1)$ vortex is given by
\begin{equation}
\Phi_\L=v_\L\diag(1,1,1),\qquad
\Phi_\R=v_\R\diag\left(f(r)e^{\i\theta},1,1\right),\qquad
A = -\i\epsilon_{ij}\frac{x^j}{2r^2}a(r)T \d x^i,
\label{eq:01vortex_ansatz}
\end{equation}
with the matrix $T=\diag(\tfrac23,-\tfrac13,-\tfrac13)$ and the VEV of $\Phi_\L$ ($\Phi_\R$) being $v_\L$ ($v_\R$), depending on the ground state in question.

In order to understand the normalization of the gauge field, we consider the split of the winding of the scalar field in eq.~\eqref{eq:10vortex_ansatz} into the global $\U(1)_{\rm B}$, $\U(1)_{\rm A}$, $\SU(3)_\L$ and $\SU(3)_\R$ with right action as well as the local $\SU(3)_{\rm C}$ with a left action:
\begin{align}
\diag\left(f(r)e^{\i\theta},1,1\right)
&=e^{\i\beta\theta T}\diag\left(f(r),1,1\right)e^{\i\beta\theta T}e^{\i\alpha\theta}e^{\i\alpha\theta},\\
\diag\left(1,1,1\right)
&=e^{\i\beta\theta T}\diag\left(1,1,1\right)e^{-\i\beta\theta T}e^{\i\alpha\theta}e^{-\i\alpha\theta}.
\end{align}
The second equation, corresponding to the right field is trivially solved by any $\alpha$ and any $\beta$, whereas the first equation leads to two equations: $\frac43\beta+2\alpha=1$ and $-\frac23\beta+2\alpha=0$, yielding $\alpha=\frac16$ and $\beta=\frac12$.
This means that the gauge part of the vortex carries the flux $\frac12T$, which indeed is the normalization of the gauge field in eq.~\eqref{eq:10vortex_ansatz}.

The boundary conditions for the two profile functions are
\beq
f(0)=0,\qquad
a(0)=0,\qquad
f(\infty)=1,\qquad
a(\infty)=1.
\eeq
A suitable initial guess for the profile functions takes the perturbative masses into account
\beq
f_{\rm guess}(r)=\tanh(m_\Phi r),\qquad
a_{\rm guess}(r)=\tanh(m_A r).
\label{eq:fa_guess}
\eeq

For the initial state, we prepare a $(1,0)$ and a $(0,1)$ vortex with large enough separation that we can assume the following Abrikosov Ansatz
\begin{align}
\Phi_\L&=v_\L\diag\left(f(r_\L)e^{\i\theta_\L},1,1\right),\qquad
\Phi_\R=v_\R\diag\left(f(r_\R)e^{\i\theta_\R},1,1\right),\non
A&=-\epsilon_{ij}\left(\frac{x_\L^j}{2r_\L^2}a(r_\L)+\frac{x_\R^j}{2r_\R^2}a(r_\R)\right)T\d x^i,
\end{align}
with $x+L+\i y=r_\L e^{\i\theta_\L}$ and 
$x-L+\i y=r_\R e^{\i\theta_\R}$ being two radial coordinates centered at the left and the right-hand side vortex, respectively.

The terms with the coefficients $\gamma_{1,2,3}$ give 
direct couplings between  the left $\Phi_\L$ and right $\Phi_\R$ condensates.
When all $\gamma_{1,2,3}$ are turned off, 
chiral vortices $(1,0)$ and $(0,1)$ are deconfined; 
they are attached by no domain walls. 
If we turn on at least one of $\gamma_{1,2,3}$, 
they are attached by one or two domain walls \cite{Eto:2021nle}, as axion strings. 
This can be confirmed by substituting the Ans\"atze in 
eq.~(\ref{eq:10vortex_ansatz}) or (\ref{eq:01vortex_ansatz}) into the potential term  
and by evaluating it at a large circle encircling a vortex. 
We then obtain a sine-Gordon model (when only one of $\gamma_{1,2,3} \neq 0$), a double sine-Gordon model 
(when $\gamma_{1,2} \neq 0$ and
$\gamma_3 = 0$) and so on. The effective sine-Gordon models count the number of domain walls attached to the vortex that we are considering.\footnote{
In the two-Higgs doublet models 
as an extension of the Standard Model, a single non-Abelian vortex is attached by one or two domain walls depending on the parameters as shown  by the same analysis \cite{Eto:2018hhg,Eto:2018tnk}. 
This model also admits  
a molecule of two non-Abelian vortex strings \cite{Eto:2021dca}.
}
In particular, in the case of $\gamma_{1,2} \neq 0$ and
$\gamma_3 = 0$, the domain walls are 
non-Abelian sine-Gordon solitons 
carrying ${\mathbb C}P^2$ moduli \cite{Nitta:2014rxa,Eto:2015uqa}.\footnote{The U$(N)$ non-Abelian sine-Gordon model appears also in the U$(N)$ chiral Lagrangian  \cite{Eto:2021gyy} and on a Josephson junction of 
two color superconductors \cite{Nitta:2015mma,Nitta:2015mxa,Nitta:2022ahj}, and 
a sine-Gordon soliton can host SU$(N)$ Skyrmions as ${\mathbb C}P^{N-1}$ lumps \cite{Eto:2015uqa,Eto:2023tuu,Nitta:2022ahj}.
}
When  
a chiral vortex is attached to a non-Abelian sine-Gordon soliton 
their ${\mathbb C}P^2$ moduli match. 
The term with $\gamma_3 \neq 0$ somehow ``Abelianizes'' the ${\mathbb C}P^2$ moduli.

\subsection{Diagonal matrices}

In ref.~\cite{Eto:2021nle} it was shown that $\CP^2$ moduli attract energetically.
Under this assumption, choosing one vortex, say the left vortex in $\Phi_\L$ to be on diagonal form (without loss of generality), the right vortex in $\Phi_\R$ will align with the other and hence also be on diagonal form.
Although we do not limit our simulations to diagonal matrices, a faster version that operates only using diagonal matrices can speed up the numerical investigations.
For this reason we give the following Lemma.
\begin{lemma}
The variational equations \eqref{eq:eom_PhiL}-\eqref{eq:eom_A} remain diagonal matrices when sourced by initial conditions that consist of diagonal matrices.
\end{lemma}
\emph{Proof}:
In order to facilitate the proof, we assume that the numerical method updates the fields by adding a constant times the equation of motion for that field to itself at every step.
This is the case for the method given in sec.~\ref{sec:numerical_method}.
It remains to check that the eom of eqs.~\eqref{eq:eom_PhiL}-\eqref{eq:eom_A} are diagonal matrices if the $\Phi_\L$, $\Phi_\R$, $A$ all are.
Since the trace preserves the diagonal structure and the variational equations consist of products of diagonal matrices, only the symbols $\Xi_{\L,\R}$ need to be checked.
An explicit calculation reveals that $\Phi_{\L,\R}$ being diagonal matrices reduces the variation of the determinant to
\beq
(\Xi_\L)_{\alpha a}=\frac12\sum_{b,c=1}^3\epsilon_{\alpha bc}\epsilon_{abc}(\Phi_\R)_{\alpha\alpha}(\Phi_\L^\dag\Phi_\R)_{bb}(\Phi_\L^\dag\Phi_\R)_{cc},
\eeq
which vanishes when $\alpha\neq a$ and similarly for $\Xi_\R$.\hfill$\square$

\subsection{Numerical method}\label{sec:numerical_method}

We will utilize the numerical method sometimes called arrested Newton flow, which similarly to relativistic dynamics accelerates towards the nearest local minimum of the energy functional. 
Unlike relativistic dynamics, the arrested part of the method is a continuous monitoring of the static energy (potential energy including field gradients) which sets the kinetic energy of the flow to zero once the energy increases.
Specifically, we solve the equations
\beq
\p_\tau^2\Phi_\L=-\eom_{\Phi_\L},\qquad
\p_\tau^2\Phi_\R=-\eom_{\Phi_\R},\qquad
\p_\tau^2A=-\eom_{A},
\eeq
with an initial condition given in sec.~\ref{sec:ansatze} with $L$ typically set to $L:=4$.
$\tau$ is not real time, but simply a parametrization of the flow.
At every step of the flow, we compute the energy $E$ of eq.~\eqref{eq:E} and compare it with the energy of the previous step.
If the energy has increased, we set $\p_\tau\Phi_\L=\p_\tau\Phi_\R=\p_\tau A=0$.

The numerical computations are performed on square lattices with $1024^2$ lattice sites and the discrete derivatives are approximated using a 5-point stencil and a 4th-order numerical derivative. The lattice spacing is typically $0.0391$.

\subsection{Numerical results}

We will now perform numerical computations with the method described
in sec.~\ref{sec:numerical_method} and the initial conditions given in
sec.~\ref{sec:ansatze}.
The main results are for the $(1,0)$ + $(0,1)$ vortices, each with
winding in the 11 components of both the left and right
scalar fields.
Due to the large parameter space of the model (9-dimensional parameter
space) and large number of possible initial conditions (relative
$\CP^2$ coordinate), we cannot claim that we have exhausted all
possibilities, but we get a general picture of how the set of chiral
vortices behave.
Also for simplicity, we will henceforth set $g:=1$ and $m:=\sqrt{2}$. 

\begin{figure}[!htp]
\centering
\mbox{\includegraphics[width=0.32\linewidth]{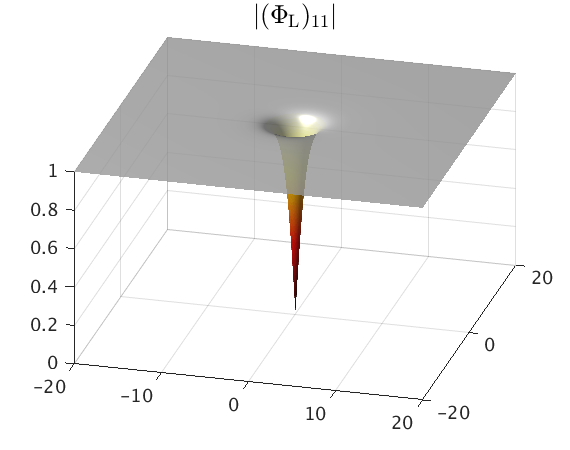}
\includegraphics[width=0.32\linewidth]{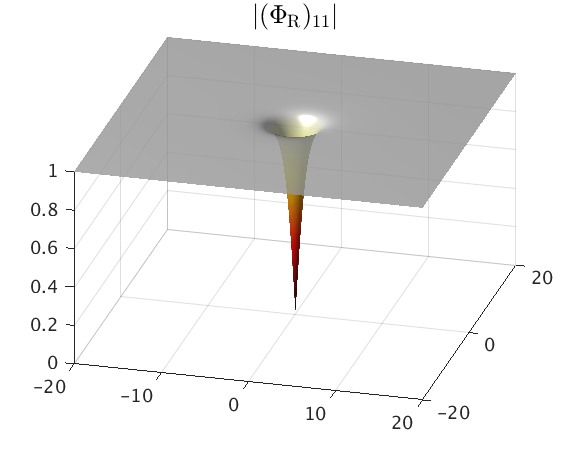}
\includegraphics[width=0.32\linewidth]{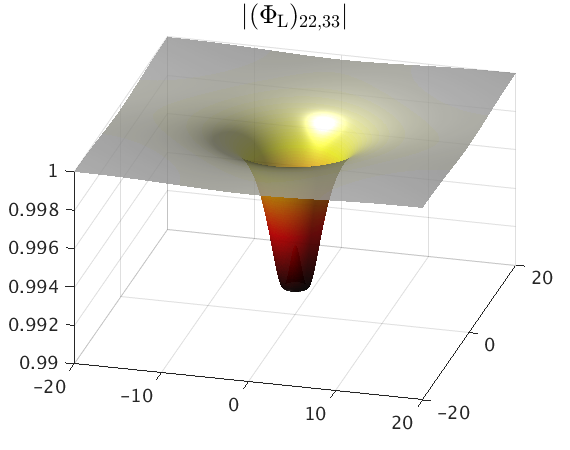}}
\mbox{\includegraphics[width=0.32\linewidth]{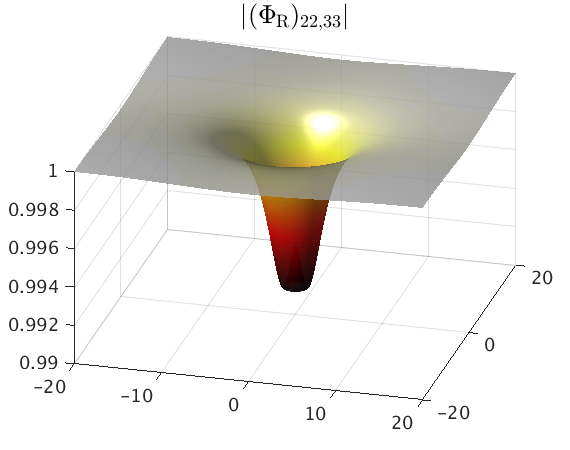}
\includegraphics[width=0.32\linewidth]{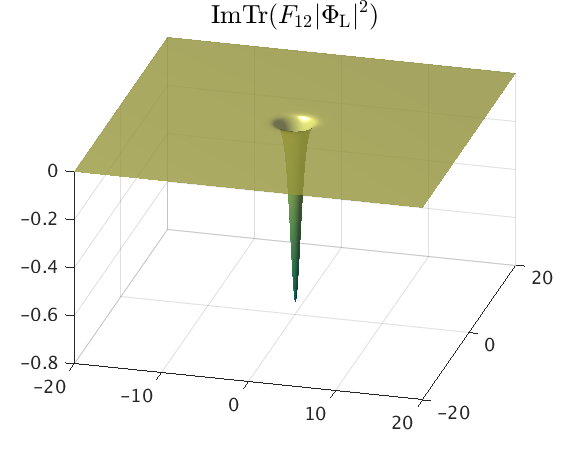}
\includegraphics[width=0.32\linewidth]{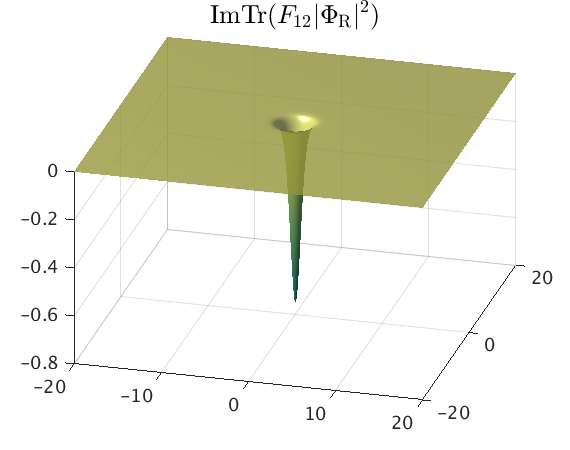}}
\mbox{\includegraphics[width=0.32\linewidth]{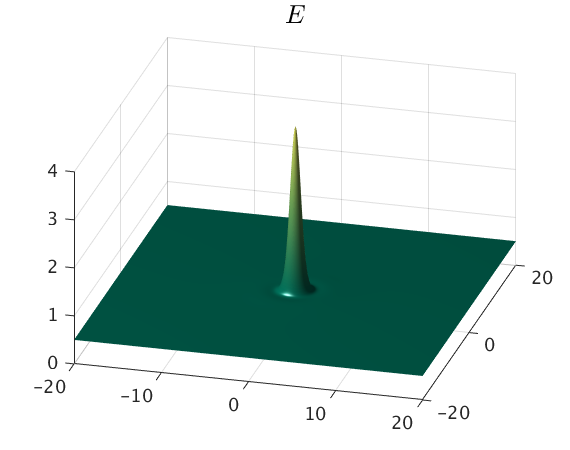}}
\caption{``Regular'' 
(axisymmetric) non-Abelian vortex, i.e.~no $\gamma$ terms turned on.
In this figure $g=1$, $m=\sqrt{2}$, $\lambda_{1,2,3,4}=(2,0,0,0)$, $\gamma_{1,2,3}=(0,0,0)$. This vortex configuration has winding number $(1,1)$, the left and
right fields' vortices are coincident and the energy is axially
symmetric ($\sfe=0$). }
\label{fig:regular}
\end{figure}
Starting with the most simplistic choice of the potential parameters,
we turn on only $\lambda_1$ (here and henceforth, the parameters
$\lambda_1$ through $\lambda_4$ and $\gamma_1$ through $\gamma_3$
vanish when not turned on), i.e.~we will set $\lambda_1:=2$.
We are surprised to see that the two vortices attract and form a
coincident vortex with total winding 
$(1,1)$, see fig.~\ref{fig:regular}.
The left and right fields' vortices are completely coincident and the
energy has eccentricity zero, where we define the eccentricity
as\footnote{We assume here that the configuration is elliptic with the
major axis along the $x$-direction; otherwise the inverse of the
fraction in the square root needs to be used.}
\beq
\sfe = \sqrt{1-\frac{\int_M \star(x-x_{\rm CM})^2\calE}{\int_M \star(y-y_{\rm CM})^2\calE}},
\eeq
where center of mass is defined as $(x_{\rm CM},y_{\rm CM})=\int_M\star (x,y)\calE$,
the energy density is defined by $E=\int_M\star\calE$ and $E$ is
the energy given in eq.~\eqref{eq:E}.
Eccentricity zero corresponds to an axially symmetric energy
configuration.

In the figure, we can clearly see the vortex ``zero'' in the 11
component of both the left and right scalar fields.
Due to the gauge field being $\SU(3)$ and hence traceless, the gauge
field must turn on the 22 and 33 components, which in turn induce
nontrivial behavior in the 22 and 33 component of both the scalar
fields.
The ``dip'' in the 22 and 33 components of the scalar fields is,
however, quite mild.
The gauge field being $\SU(3)$ also implies that its field strength
$F_{12}$ is traceless. The non-Abelian part of the field strength is,
however, not gauge invariant.
We thus display a gauge invariant quantity constructed out of the
non-Abelian field strength as well as the two scalar fields
\beq
{\rm Im}\big[\Tr[F_{12}\Phi_\L\Phi_\L^\dag]\big], \qquad
{\rm Im}\big[\Tr[F_{12}\Phi_\R\Phi_\R^\dag]\big],
\eeq
where taking the imaginary part is simply due to the convention of
using anti-Hermitian gauge fields.
The reason for multiplying by $\Phi_\L\Phi_\L^\dag$ is that it is a
matrix in color indices (i.e.~with the flavors traced over). 

\begin{figure}[!htp]
\centering
\mbox{\includegraphics[width=0.32\linewidth]{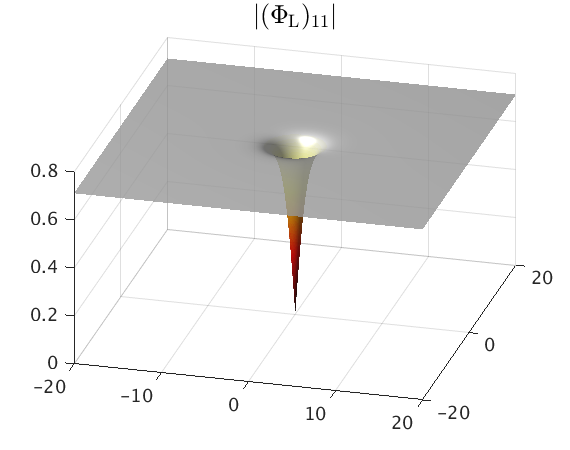}
\includegraphics[width=0.32\linewidth]{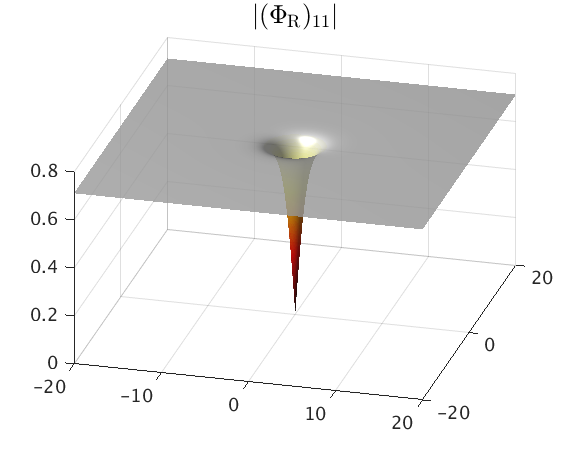}
\includegraphics[width=0.32\linewidth]{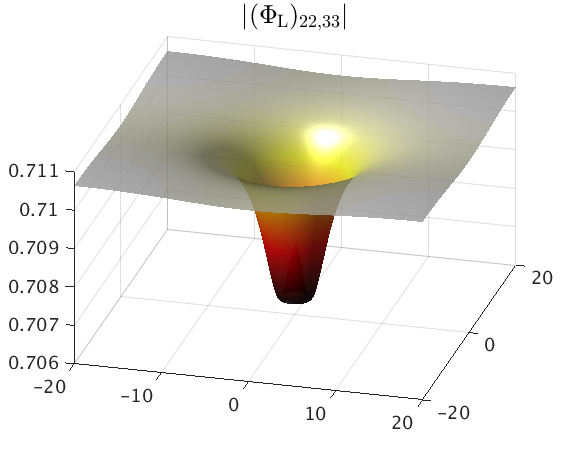}}
\mbox{\includegraphics[width=0.32\linewidth]{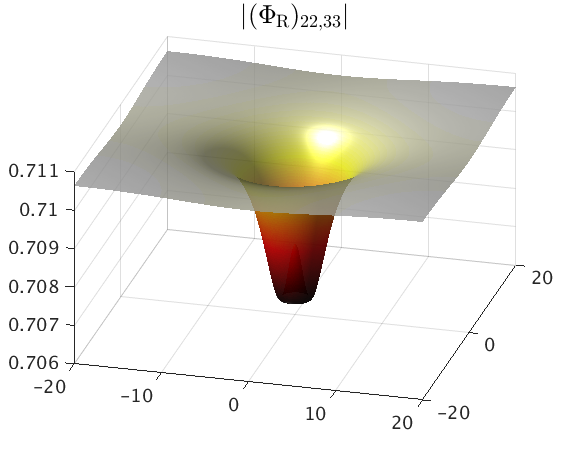}
\includegraphics[width=0.32\linewidth]{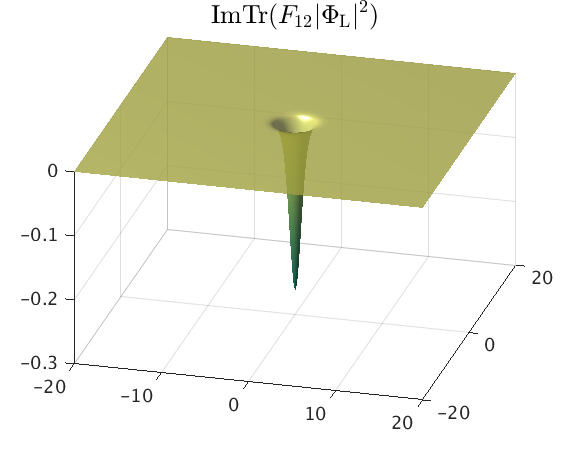}
\includegraphics[width=0.32\linewidth]{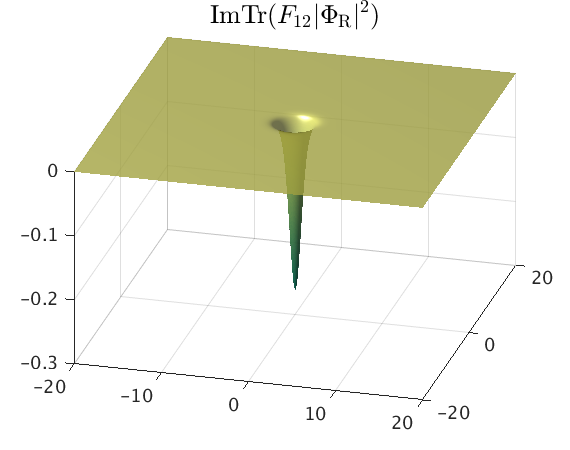}}
\mbox{\includegraphics[width=0.32\linewidth]{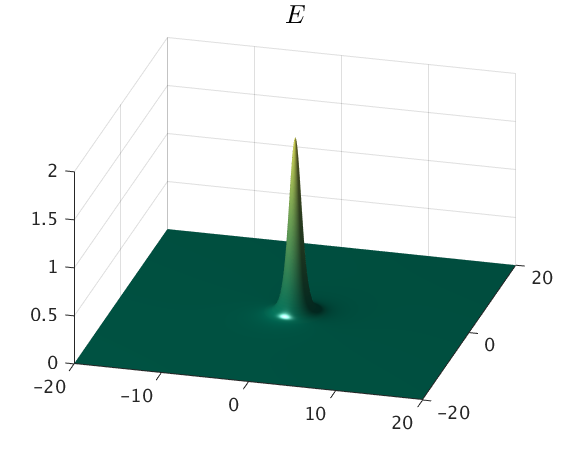}
\includegraphics[width=0.32\linewidth]{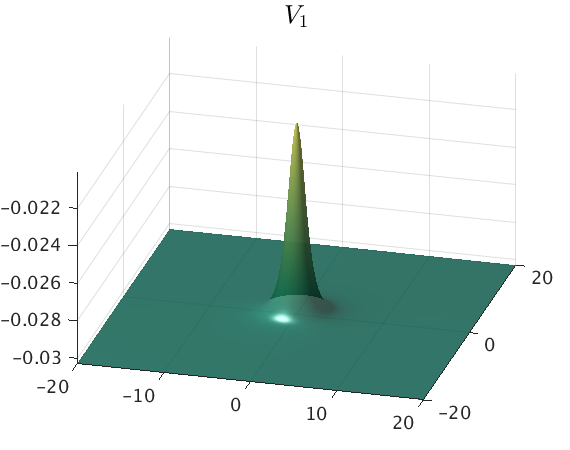}}
\caption{Confined chiral vortices.
In this figure $g=1$, $m=\sqrt{2}$, $\lambda_{1,2,3,4}=(3,0,0,1)$, $\gamma_{1,2,3}=(-0.01,0,0)$. This vortex configuration has winding number $(1,1)$, the left and
right fields' vortices are coincident and the energy is axially
symmetric ($\sfe=0$). }
\label{fig:confined}
\end{figure}
In fig.~\ref{fig:confined}, we turn on the $\lambda_4$ and also the
Josephson term, i.e.~$\gamma_1$.
$\lambda_4$ does not have much impact on the vortex configuration,
as long as it is smaller than $\lambda_1$.
Taking it larger than $\lambda_1$ changes the ground state structure, but as
we will see shortly, once it is of the same magnitude as $\lambda_1$,
it will have an impact on the outcome.

By the logic that the Josephson term $(\gamma_1)$ attaches one domain
wall connecting the left and right chiral vortices, whereas the
$\gamma_2$ term attaches two domain walls between the pair of vortices
\cite{Eto:2021nle}, we can predict that the tension of the
Josephson wall will give rise to further attraction.
In the example given in fig.~\ref{fig:confined}, this is indeed the
case, but since the vortices already want to be coincident, nothing
much can be seen from turning on the Josephson term
($\gamma_1\neq0$). 

\begin{figure}[!htp]
\centering
\mbox{\includegraphics[width=0.32\linewidth]{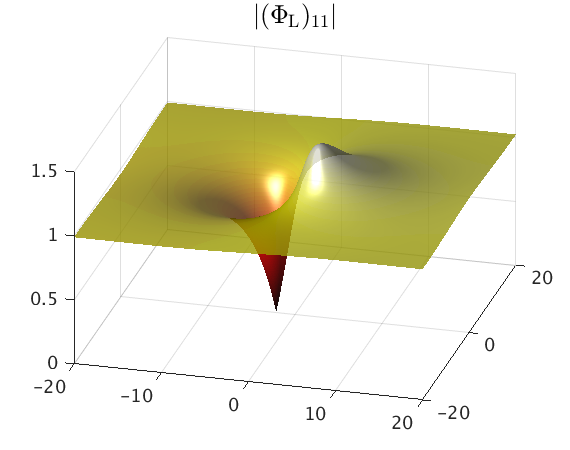}
\includegraphics[width=0.32\linewidth]{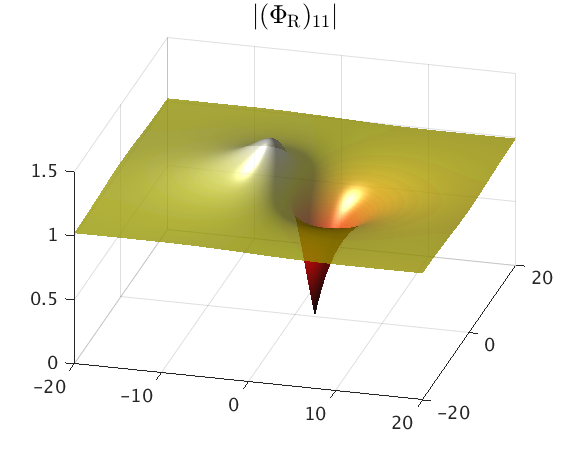}
\includegraphics[width=0.32\linewidth]{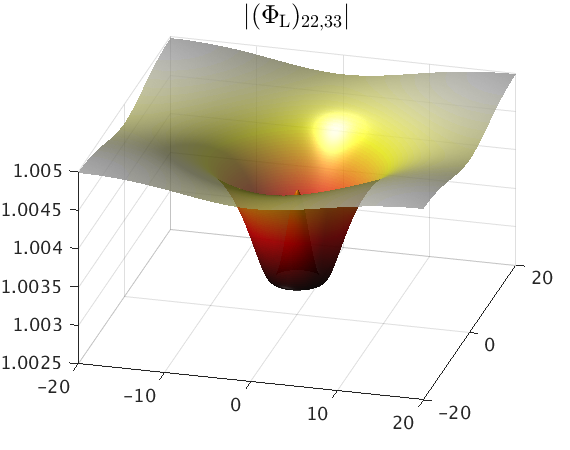}}
\mbox{\includegraphics[width=0.32\linewidth]{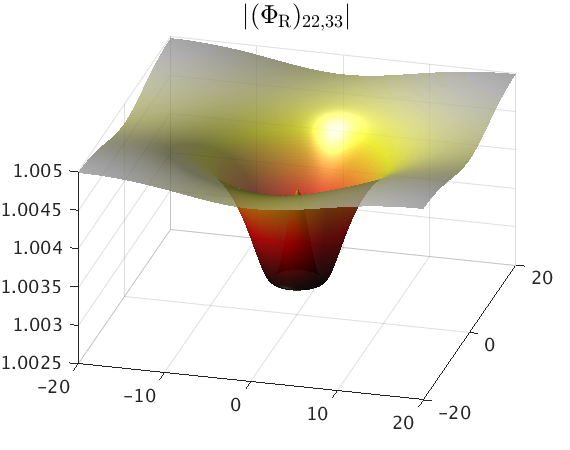}
\includegraphics[width=0.32\linewidth]{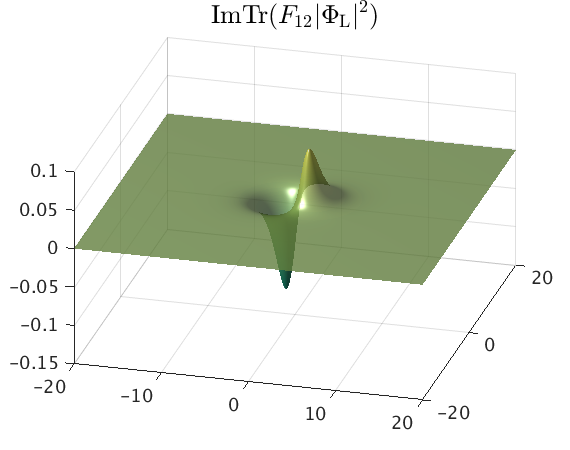}
\includegraphics[width=0.32\linewidth]{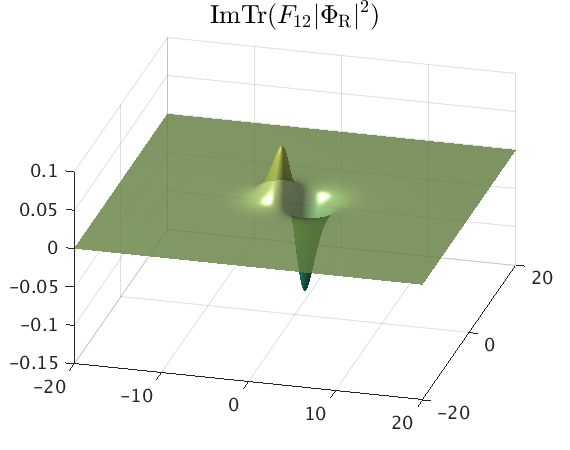}}
\mbox{\includegraphics[width=0.32\linewidth]{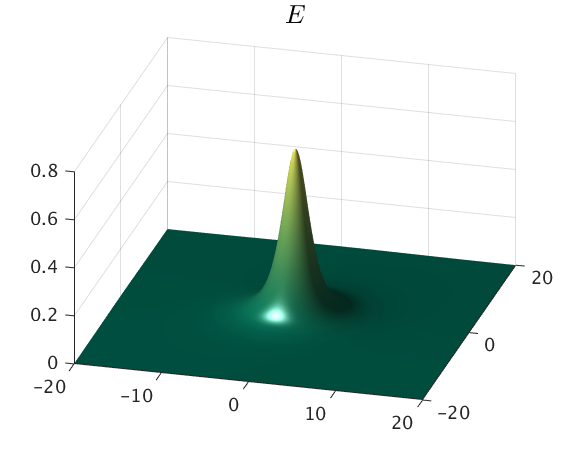}
\includegraphics[width=0.32\linewidth]{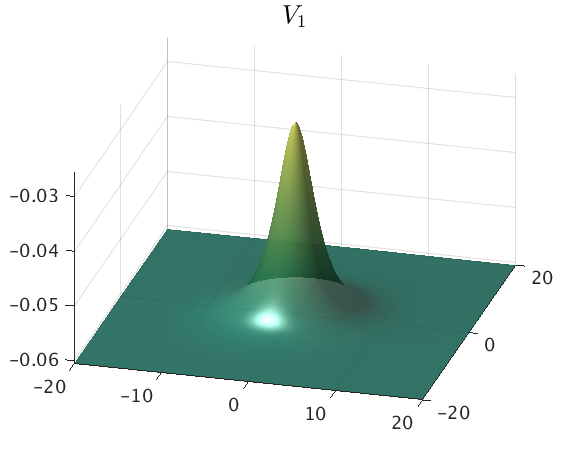}}
\caption{Dipolar chiral vortices.
In this figure $g=1$, $m=\sqrt{2}$, $\lambda_{1,2,3,4}=(1,0,0,1)$, $\gamma_{1,2,3}=(-0.01,0,0)$. This vortex configuration has winding number $(1,1)$, the left and
right fields' vortices form a dipole but the energy is axially
symmetric ($\sfe=0$). The scalar fields break the axial symmetry though. }
\label{fig:pill}
\end{figure}

In order to provoke some nontrivial structure out of the composite
chiral vortices that like to be sitting in a coincident bound state,
we leave the Josephson term on $\gamma_1\neq0$, but lower $\lambda_1$
to the same level of $\lambda_4$, see fig.~\ref{fig:pill}.
This gives rise to a repulsion of the left and right chiral vortices,
that are, however, still confined by the Josephson wall.
Interestingly, although the scalar field clearly have distinct zeros
(non-coincident) the energy is nevertheless axially symmetric.
In that sense, just like a magnet, the bound state is dipolar with the
two zeros forming the two poles of the state, which however remains
axially symmetric in terms of the energy density. The eccentricity thus
remains vanishing.
In fig.~\ref{fig:pill} we also display $V_1$ which is the $\gamma_1$
part of $V$ (see eq.~\eqref{eq:V}). 

\begin{figure}[!htp]
\centering
\mbox{\includegraphics[width=0.32\linewidth]{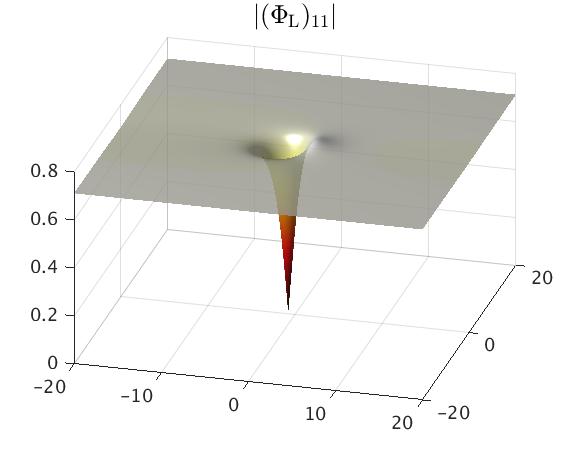}
\includegraphics[width=0.32\linewidth]{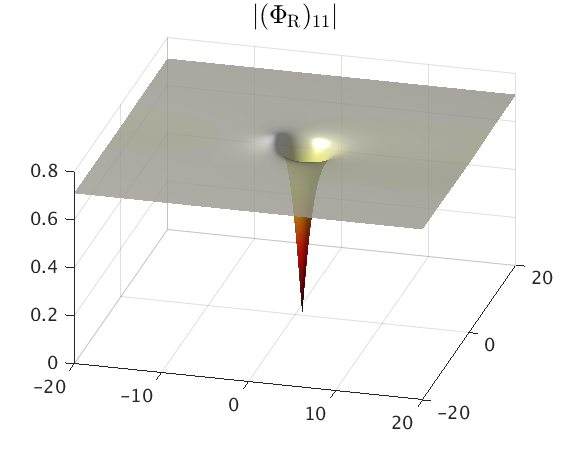}
\includegraphics[width=0.32\linewidth]{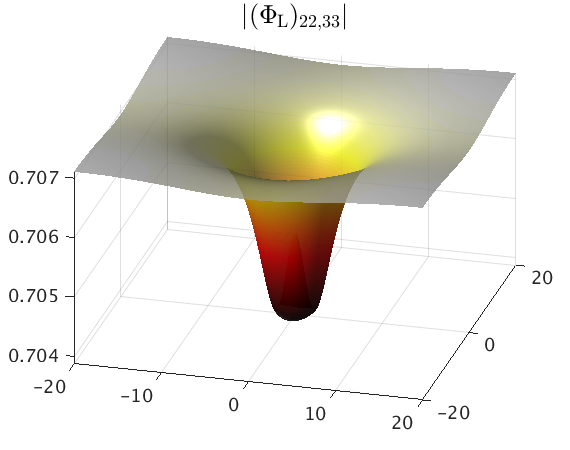}}
\mbox{\includegraphics[width=0.32\linewidth]{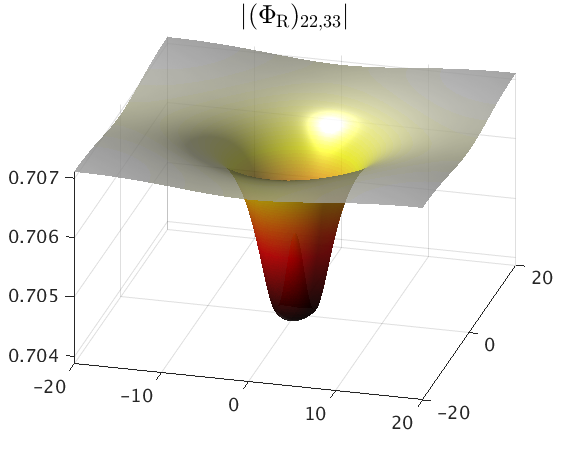}
\includegraphics[width=0.32\linewidth]{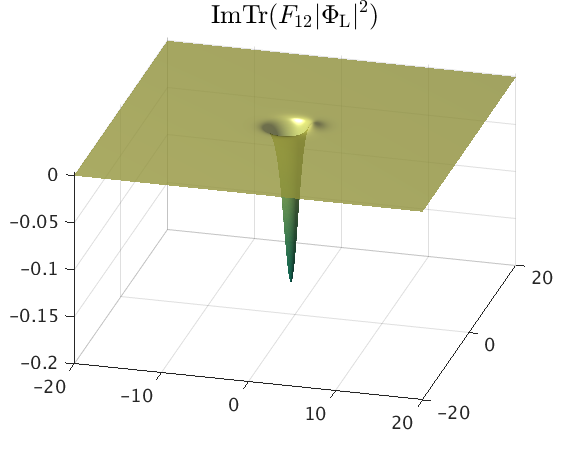}
\includegraphics[width=0.32\linewidth]{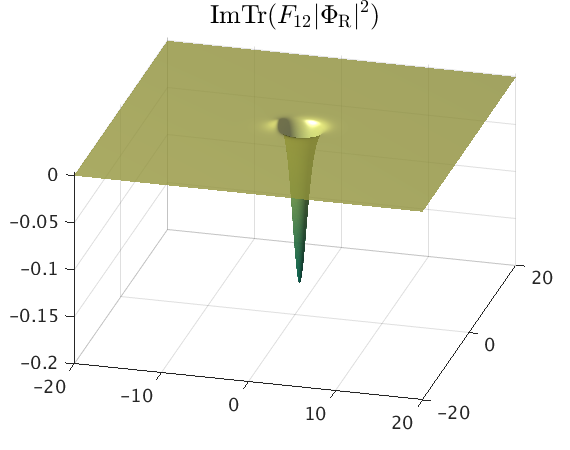}}
\mbox{\includegraphics[width=0.32\linewidth]{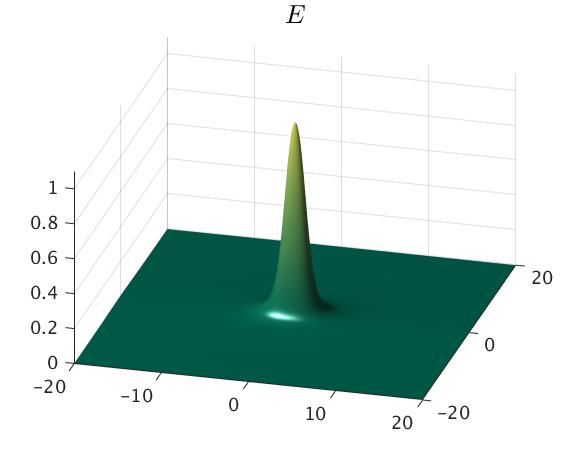}
\includegraphics[width=0.32\linewidth]{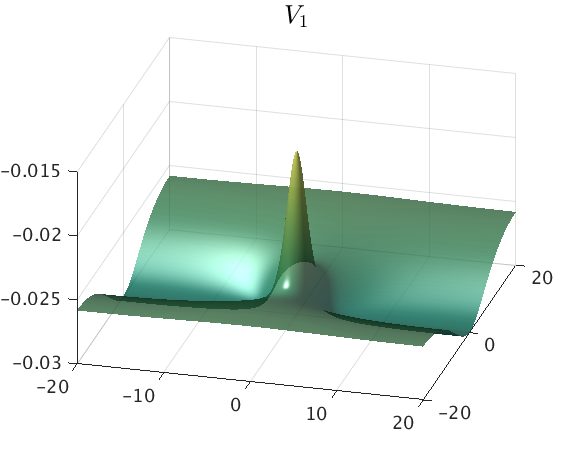}
\includegraphics[width=0.32\linewidth]{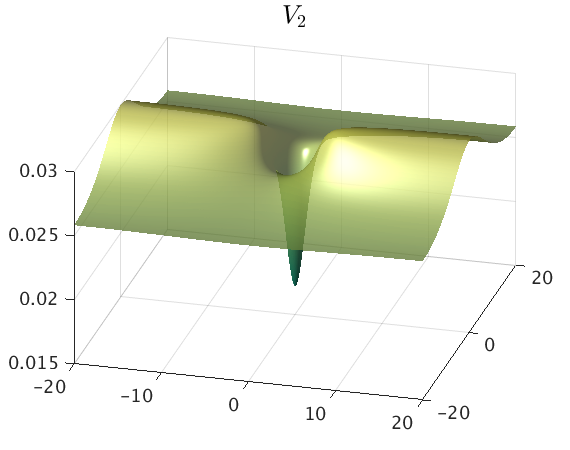}}
\caption{Chiral vortices with both Josephson and Josephson-squared terms turned on.
In this figure $g=1$, $m=\sqrt{2}$, $\lambda_{1,2,3,4}=(3,0,0,1)$, $\gamma_{1,2,3}=(-0.01,0.01,0)$. This vortex configuration has winding number $(1,1)$, the left and
right fields' vortices form a dipole and the energy density is elliptic
with eccentricity $\sfe=0.71$.  The vortex dipolar molecule lives on
an infinite Josephson wall, which however has vanishing energy
density. }
\label{fig:balance}
\end{figure}

The way the Josephson term ($\gamma_1$) and its squared generalization
($\gamma_2$) connects to two vortices depends on the sign of the
coefficients (or alternatively also on how the winding of the vortices
is chosen; for example with $\gamma_2=\gamma_3=0$ the configurations
are symmetric under $\gamma_1\to-\gamma_1$ and $\Phi_\R\to-\Phi_R$). 
In particular, the Josephson walls can form infinite domain walls.
In fact, the non-axially symmetric vortex configuration shown in
fig.~\ref{fig:balance} is created by having each vortex (i.e.~left and
right vortices) have a $\gamma_1$-wall that tends off to infinity,
but meticulously choosing the $\gamma_2$ term with the exact same
magnitude and opposite sign. This creates a small molecular (dipolar)
bound state of left and right vortices, that are connected with one
$\gamma_1$-wall, but have the other two $\gamma_2$-walls tending
off to spatial infinity (recall that the $\gamma_2$ term attaches two
walls to each vortex).
Because of the fine tuned magnitude and opposite sign of the couplings
$\gamma_2=-\gamma_1$, the energy of the wall tending off to infinity
cancels out exactly (see $V_1$ and $V_2$ in fig.~\ref{fig:balance}),
leaving behind a confined albeit non-axially symmetric dipolar
molecule of chiral vortices.\footnote{
This resembles a domain-wall bimeron in chiral magnets~\cite{Amari:2024jxx}.
}

\begin{figure}[!htp]
\centering
\mbox{\includegraphics[width=0.32\linewidth]{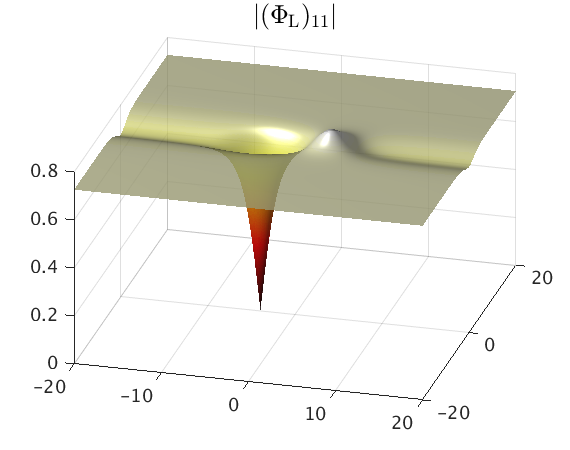}
\includegraphics[width=0.32\linewidth]{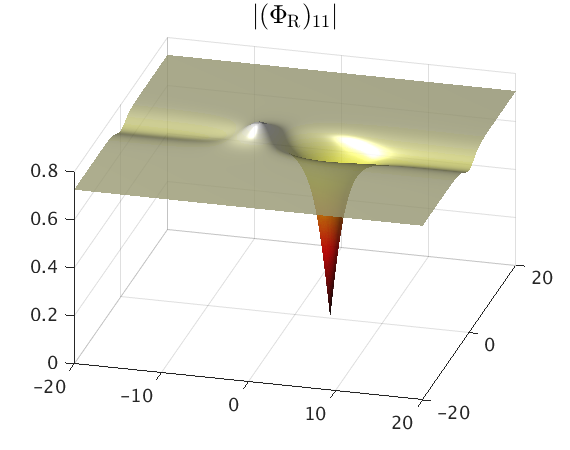}
\includegraphics[width=0.32\linewidth]{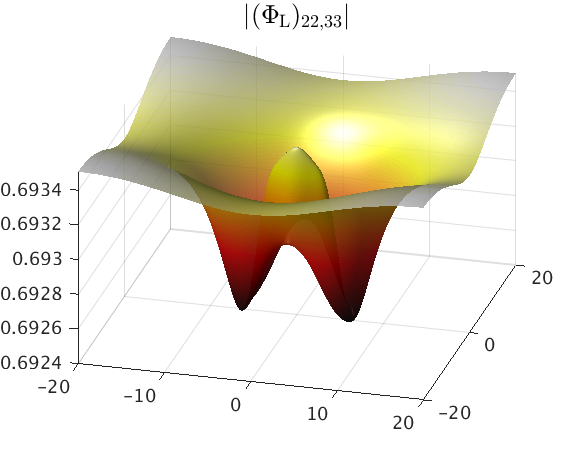}}
\mbox{\includegraphics[width=0.32\linewidth]{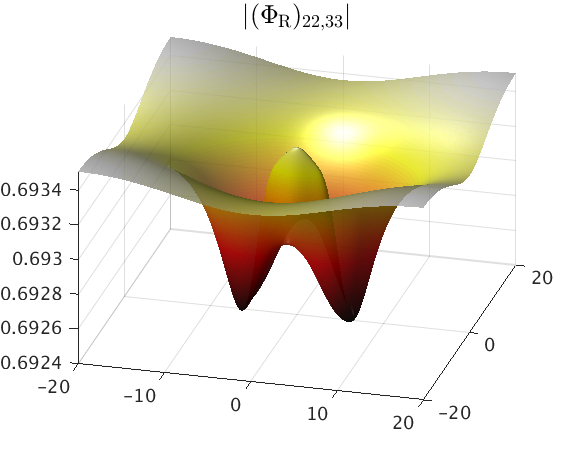}
\includegraphics[width=0.32\linewidth]{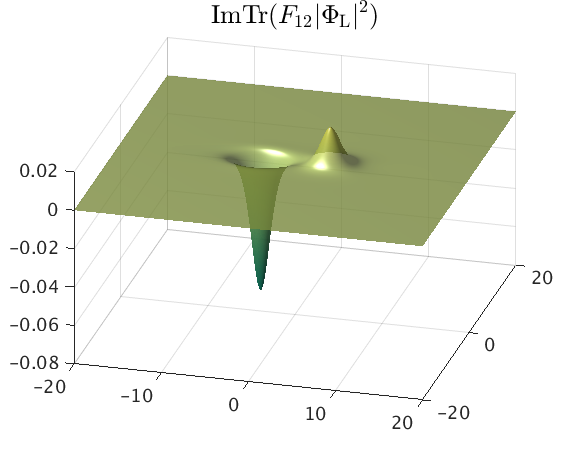}
\includegraphics[width=0.32\linewidth]{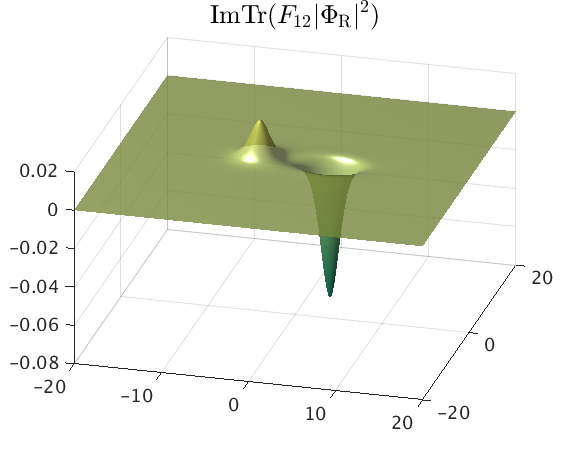}}
\mbox{\includegraphics[width=0.32\linewidth]{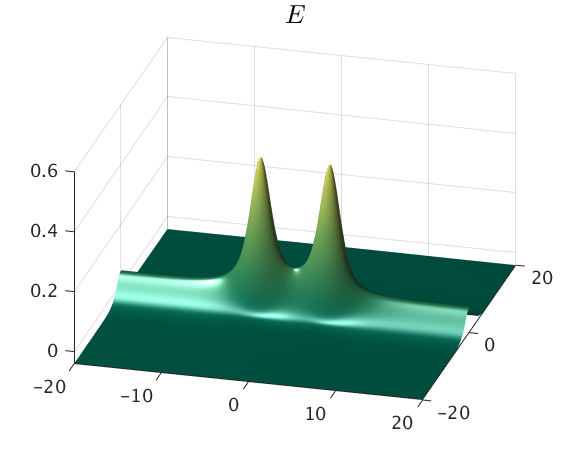}
\includegraphics[width=0.32\linewidth]{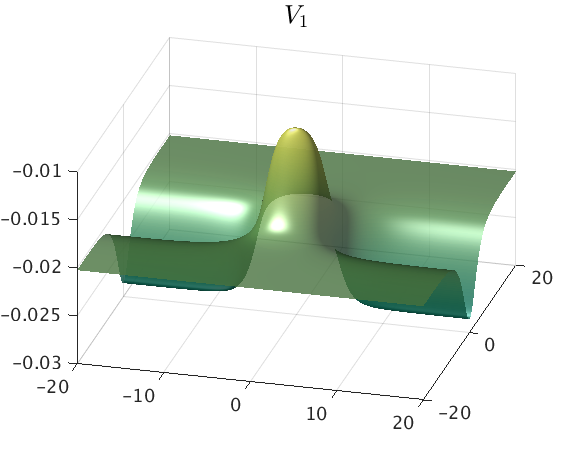}
\includegraphics[width=0.32\linewidth]{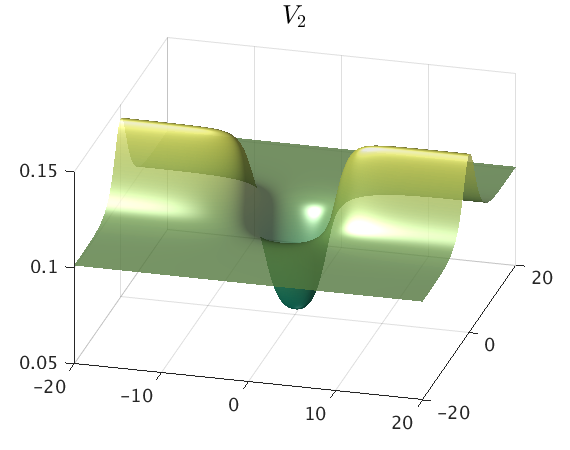}}
\caption{Chiral vortices as beats on a wall (Josephson double domain wall from the Josephson-squared term).
In this figure $g=1$, $m=\sqrt{2}$, $\lambda_{1,2,3,4}=(3,0,0,1)$, $\gamma_{1,2,3}=(-0.01,0.05,0)$. This vortex configuration has winding number $(1,1)$, the left and
right fields' vortices form a dipole and the energy density is elliptic
with eccentricity $\sfe=0.95$ (computed on the domain of the
figure). The vortex dipolar molecule lives on an infinite Josephson
wall with finite tension. }
\label{fig:beats}
\end{figure}

Now if we do not fine tune the cancellation between the Josephson term
and its squared counterpart, we will create a domain wall with finite
energy, see fig.~\ref{fig:beats}.
Increasing $\gamma_2$ creates an infinite double Josephson wall from
each vortex. They are repelled slightly by the $\gamma_1$ term and
confined by the $\gamma_2$ term, which however also gives an infinite
contribution to the energy (if space is taken to be $\mathbb{R}^2$).
This chiral vortex configuration is truly a molecule of $(1,0)$ and
$(0,1)$ vortices with a visible separation (non-coincidence) in both
the field zeros and in the energy density.

\begin{figure}[!htp]
\centering
\mbox{\includegraphics[width=0.32\linewidth]{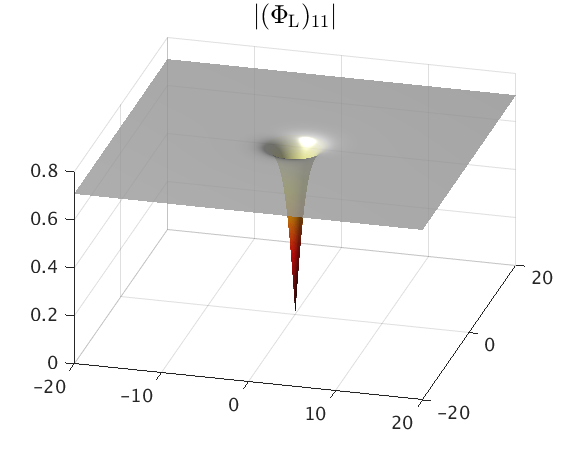}
\includegraphics[width=0.32\linewidth]{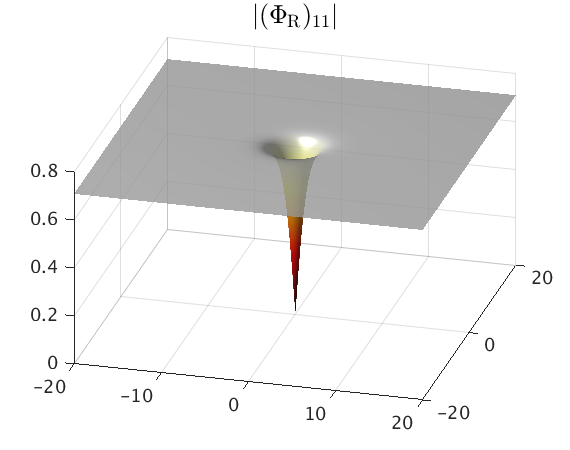}
\includegraphics[width=0.32\linewidth]{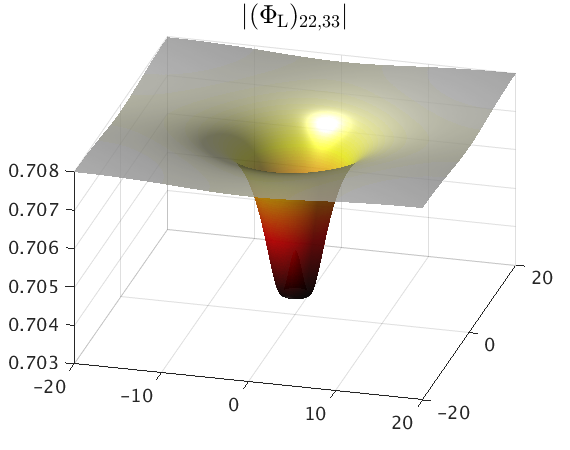}}
\mbox{\includegraphics[width=0.32\linewidth]{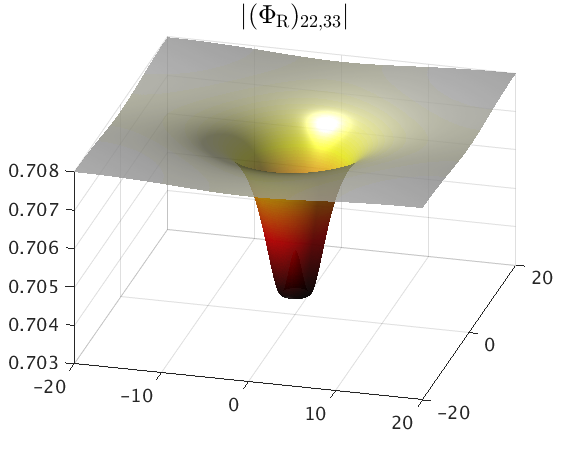}
\includegraphics[width=0.32\linewidth]{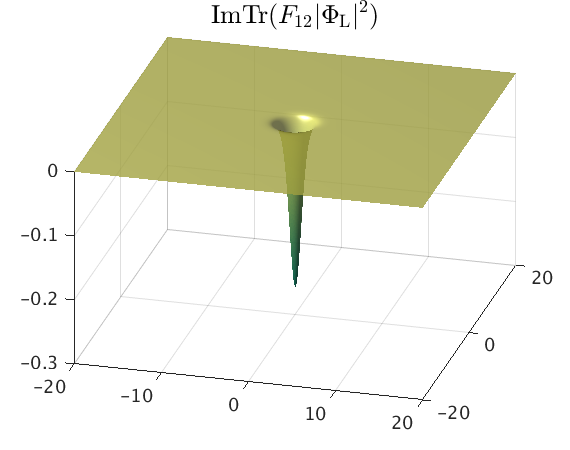}
\includegraphics[width=0.32\linewidth]{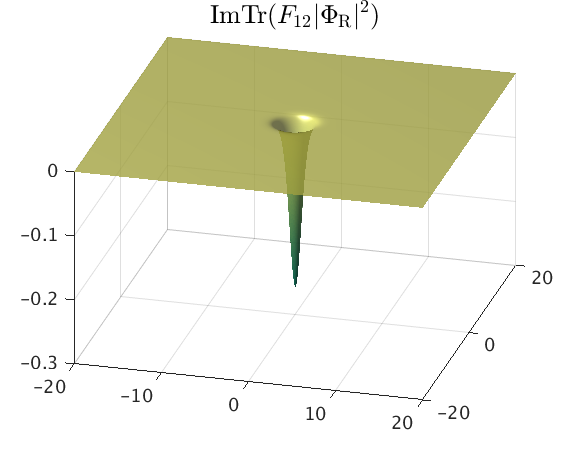}}
\mbox{\includegraphics[width=0.32\linewidth]{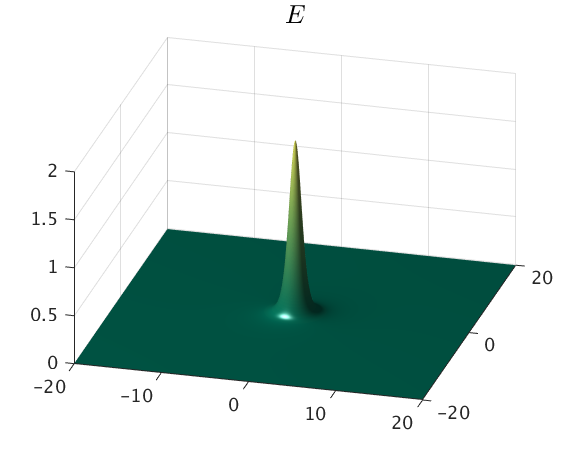}
\includegraphics[width=0.32\linewidth]{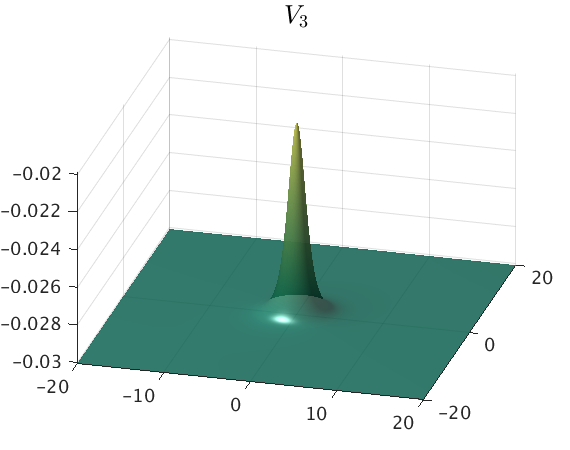}}
\caption{Confined vortices with the $\gamma_3$ term.
In this figure $g=1$, $m=\sqrt{2}$, $\lambda_{1,2,3,4}=(3,0,0,1)$, $\gamma_{1,2,3}=(0,0,-0.01)$. This vortex configuration has winding number $(1,1)$, the left and
right fields' vortices are coincident and the energy is axially
symmetric ($\sfe=0$). }
\label{fig:det}
\end{figure}

There are many other configurations that are very similar in nature to
the examples we have selected above.
As long as the ground state conditions allow for the same fully broken and
symmetric (up to a phase or sign) ground state, many sets of coupling values
give rise to very similar types of chiral vortex bound states.
In fig.~\ref{fig:det}, we give an example of changing the $\gamma_1$
term for the $\gamma_3$ term in the configuration shown in
fig.~\ref{fig:confined}, yielding a very similar result.
Another rule of thumb is that interpolating between $\lambda_1$ and
$\lambda_2$ gives rise to the same type of configurations; the same
holds true for interpolating between $\lambda_3$ and $\lambda_4$.

\begin{figure}[!htp]
\centering
\includegraphics[width=0.5\linewidth]{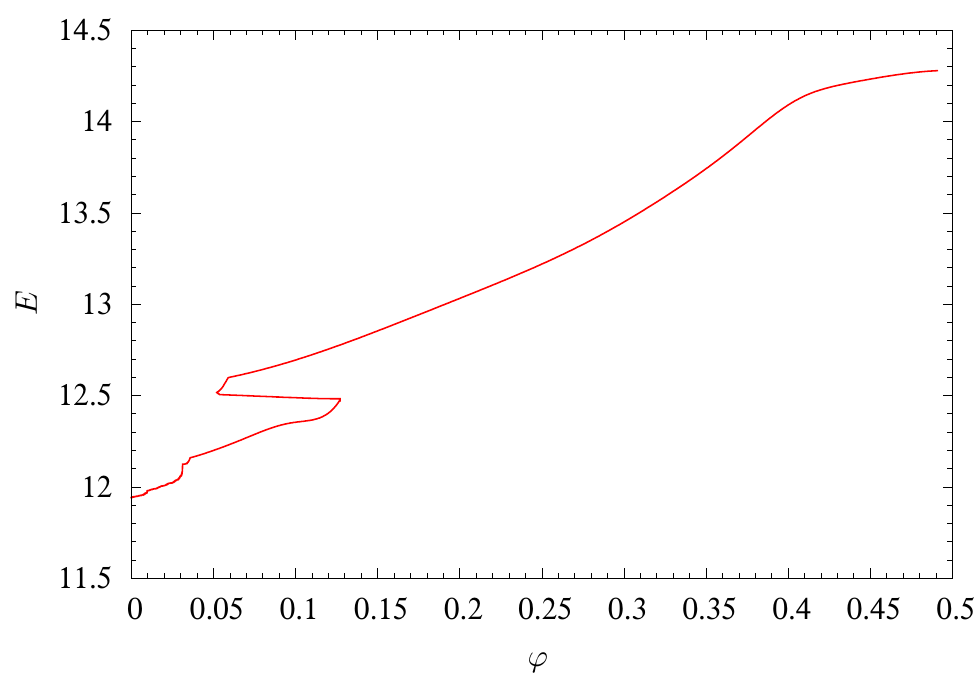}
\caption{Energy during a flow for a configuration with the left vortex
in the 11 component and the right vortex in the 22 component. The flow
minimizes the energy, eventually aligning the two vortices in
$\CP^2$. For details of the angle $\varphi$, see the text. }
\label{fig:orientation}
\end{figure}

One aspect of the huge parameter space that we left out in the above
results, is the fact that the left $(1,0)$ vortex and the right
$(0,1)$ vortex can both be rotated around in $\CP^2$.
The overall rotation (i.e.~of both vortices) makes no difference in
the energy density.
However, there is a whole $\CP^2$ space of \emph{relative}
orientations, for example fixing the left vortex to be in the 11
component, the right vortex could be rotated to other points of
$\CP^2$.
We know from ref.~\cite{Eto:2021nle} that orthogonal vortices have
larger energies than parallel ones. Let us fix the left vortex in the 11
component of $\Phi_\L$. The latter statement says that the energy is
larger if we place the right vortex in the 22 component of $\Phi_\R$
than if we place it in the 11 component.
One may contemplate what happens in between these two points of
$\CP^2$.
Using a rotation matrix of the form
\beq
U(\varphi) =
\begin{pmatrix}
  \cos\varphi & \sin\varphi & 0\\
  -\sin\varphi & \cos\varphi & 0\\
  0 & 0 & 1
\end{pmatrix},
\label{eq:Uvarphi}
\eeq
and rotating the fields as
\begin{align}
  \Phi_\L&=v_\L\diag\big(f(r_\L)e^{\i\theta_\L},1,1\big),\qquad
  \Phi_\R=v_\R U\diag\big(f(r_\R)e^{\i\theta_\R},1,1\big)U^\dag, \non
  A &= -\epsilon_{ij}\left(\frac{x_\L^j}{2r_\L^2}a(r_\L)T
  + \frac{x_\R^j}{2r_\R^2}a(r_\R) U T U^\dag\right) \d x^i,
\end{align}
with $x+L+\i y=r_\L e^{\i\theta_\L}$,
$x-L+\i y=r_\R e^{\i\theta_\R}$ as usual and $f(r)$, $a(r)$ given in
eq.~\eqref{eq:fa_guess}, we can compute the energy as a function of
$\varphi$.
We choose a generic example with the couplings chosen as in
fig.~\ref{fig:confined} and display the energy of the relative orientation rotated by $U$
of eq.~\eqref{eq:Uvarphi} in fig.~\ref{fig:orientation}.
The figure shows the energy during the flow towards the minimum of the
energy functional while tracing the orientation $\varphi$, projected
onto the $\SO(2)$ subspace of $\CP^2$.
The angle $\varphi$ is not monotonic, but this is due to projecting
onto the real subspace $\SO(2)$.
Regardless of the projection chosen, the end result is that the
vortices align their moduli to point in the same direction.
Notice, however, that they point in the same direction, but are not
pointing in the 11 direction due to the midpoint between 11 and 22
being some off-diagonal point.

\section{Conclusion and outlook}
\label{sec:summary}

In this paper, we have studied non-Abelian vortices in the CFL phase in dense 3-flavor QCD
in a non-axially symmetric setting, especially searching for
molecule-like or spatially nontrivial configurations with one vortex
in the left condensation and one in the right condensation. 
The chiral vortices can be rotated around in $\CP^2$ and by studying
the system with a left and a right vortex, this potentially leads to
quite a large space of initial conditions.
We have found, however, that the vortex moduli attract and the minimizers of
the energy functional that we found were always in the parallel
state, i.e.~both the left and the right vortex pointing in the same
direction inside $\CP^2$.
Exploring the parameter space, we have found characteristic examples of
bound states of left and right vortices, that break the axial symmetry
and finally we have found the most molecular-like state of two chiral
vortices on a chiral domain wall with two walls attached to each of
the vortices -- one of them binding the bound state and the other two
ends tending off to infinity.

It would be interesting to get a better understanding of the vortex
interactions in this model, which is quite complicated -- especially
at the nonlinear level, where the entire 9-dimensional parameter space
kicks in. At the linear level, only the mass term and the Josephson
term is probed by the scalar fields, but care must be taken for the
non-Abelian gauge field to provide accurate predictions for the vortex
interactions, which could probably be done by elaborating on the
methods developed in ref.~\cite{Speight:1996px}.

Non-axisymmetric vortex configurations found in this paper break the axial symmetry spontaneously, yielding 
a rotational Nambu-Goldstone mode. 
Such a mode, called a twiston, can propagate along the vortex line,  
as a helium-3 superfluid.
The transition between axisymmetric and non-axisymmetric configurations may affect thermal properties of a vortex lattice, which could be relevant for the core of neutron stars. 
Another direction is a possible construction of a non-Abelian vorton, i.e.~a closed vortex string with a non-trivial twist. 

In this paper,
electromagnetic interactions are neglected for simplicity. They can be taken into account in the presence of a non-Abelian vortex, 
resulting in a nontrivial potential on the ${\mathbb C}P^2$ moduli \cite{Vinci:2012mc} as well as an Aharanov-Bohm (AB) phase 
\cite{Chatterjee:2015lbf}. 
The vortex molecules in this paper also ought to be studied in the presence of  electromagnetic interactions.
The scattering of charged particles such as electrons and charged CFL pions off of a vortex should exhibit nontrivial AB phases.
In addition to the electromagnetic AB phase,
single chiral non-Abelian vortices exhibit 
non-Abelian AB phases, i.e.~when (quasi-)quarks encircle 
the vortex it picks up 
color nonsinglet AB phases 
\cite{Eto:2021nle}, 
similarly to the case of non-Abelian Alice strings 
in the 2SC + dd phase \cite{Fujimoto:2020dsa,Fujimoto:2021bes}, 
which are also confined into 
a single non-Abelian vortex
\cite{Fujimoto:2021wsr}.
The chiral non-Abelian vortex molecule  
may exhibit non-Abelian scattering if 
quarks can pass through between the two chiral vortices. 
One of the related nontrivial non-Abelian properties is the so-called topological obstruction, 
implying that generators of the unbroken symmetry in the ground state 
are not globally defined around the vortices \cite{Eto:2021nle}. This might be relevant for the topological properties of the ground state(s).

Finally, beyond the GL effective theory, it can be shown in the Bogoliubov-de Gennes equation, describing (quasi-)quark degrees of freedom, that an axisymmetric non-Abelian vortex admits  
three Majorana fermion zeromodes in its core \cite{Yasui:2010yw,Fujiwara:2011za,Chatterjee:2016ykq}. 
Such Majorana fermion zeromodes turn non-Abelian vortices into non-Abelian anyons \cite{Yasui:2010yh,Hirono:2012ad,Masaki:2023rtn}.

\subsection*{Acknowledgments}

S.~B.~G.~thanks the Outstanding Talent Program of Henan University for partial support.
The work of M.~N.~is supported in part by JSPS KAKENHI [Grant No.~JP22H01221]
and the WPI program ``Sustainability with Knotted Chiral Meta Matter (SKCM2 )'' at Hiroshima University (M.~N.).

\bibliographystyle{JHEP}
\bibliography{bib_chiral}

\providecommand{\href}[2]{#2}\begingroup\raggedright\begin{thebibliography}{10}

\bibitem{Alford:2007xm}
M.~G. Alford, A.~Schmitt, K.~Rajagopal, and T.~Sch{\"a}fer, {\it {Color
  superconductivity in dense quark matter}},  {\em Rev. Mod. Phys.} {\bf 80}
  (2008) 1455--1515, [\href{http://arxiv.org/abs/0709.4635}{{\tt
  arXiv:0709.4635}}].

\bibitem{Alford:1998mk}
M.~G. Alford, K.~Rajagopal, and F.~Wilczek, {\it {Color flavor locking and
  chiral symmetry breaking in high density QCD}},  {\em Nucl. Phys. B} {\bf
  537} (1999) 443--458, [\href{http://arxiv.org/abs/hep-ph/9804403}{{\tt
  hep-ph/9804403}}].

\bibitem{Alford:1997zt}
M.~G. Alford, K.~Rajagopal, and F.~Wilczek, {\it {QCD at finite baryon density:
  Nucleon droplets and color superconductivity}},  {\em Phys. Lett. B} {\bf
  422} (1998) 247--256, [\href{http://arxiv.org/abs/hep-ph/9711395}{{\tt
  hep-ph/9711395}}].

\bibitem{Rapp:1997zu}
R.~Rapp, T.~Sch{\"a}fer, E.~V. Shuryak, and M.~Velkovsky, {\it {Diquark Bose
  condensates in high density matter and instantons}},  {\em Phys. Rev. Lett.}
  {\bf 81} (1998) 53--56, [\href{http://arxiv.org/abs/hep-ph/9711396}{{\tt
  hep-ph/9711396}}].

\bibitem{Eto:2013hoa}
M.~Eto, Y.~Hirono, M.~Nitta, and S.~Yasui, {\it {Vortices and Other Topological
  Solitons in Dense Quark Matter}},  {\em PTEP} {\bf 2014} (2014), no.~1
  012D01, [\href{http://arxiv.org/abs/1308.1535}{{\tt arXiv:1308.1535}}].

\bibitem{Balachandran:2005ev}
A.~Balachandran, S.~Digal, and T.~Matsuura, {\it {Semi-superfluid strings in
  high density QCD}},  {\em Phys. Rev. D} {\bf 73} (2006) 074009,
  [\href{http://arxiv.org/abs/hep-ph/0509276}{{\tt hep-ph/0509276}}].

\bibitem{Nakano:2007dr}
E.~Nakano, M.~Nitta, and T.~Matsuura, {\it {Non-Abelian strings in high density
  QCD: Zero modes and interactions}},  {\em Phys. Rev. D} {\bf 78} (2008)
  045002, [\href{http://arxiv.org/abs/0708.4096}{{\tt arXiv:0708.4096}}].

\bibitem{Nakano:2008dc}
E.~Nakano, M.~Nitta, and T.~Matsuura, {\it {Non-Abelian Strings in Hot or Dense
  QCD}},  {\em Prog. Theor. Phys. Suppl.} {\bf 174} (2008) 254--257,
  [\href{http://arxiv.org/abs/0805.4539}{{\tt arXiv:0805.4539}}].

\bibitem{Eto:2009kg}
M.~Eto and M.~Nitta, {\it {Color Magnetic Flux Tubes in Dense QCD}},  {\em
  Phys. Rev. D} {\bf 80} (2009) 125007,
  [\href{http://arxiv.org/abs/0907.1278}{{\tt arXiv:0907.1278}}].

\bibitem{Eto:2009bh}
M.~Eto, E.~Nakano, and M.~Nitta, {\it {Effective world-sheet theory of color
  magnetic flux tubes in dense QCD}},  {\em Phys. Rev. D} {\bf 80} (2009)
  125011, [\href{http://arxiv.org/abs/0908.4470}{{\tt arXiv:0908.4470}}].

\bibitem{Eto:2009tr}
M.~Eto, M.~Nitta, and N.~Yamamoto, {\it {Instabilities of Non-Abelian Vortices
  in Dense QCD}},  {\em Phys. Rev. Lett.} {\bf 104} (2010) 161601,
  [\href{http://arxiv.org/abs/0912.1352}{{\tt arXiv:0912.1352}}].

\bibitem{Hanany:2003hp}
A.~Hanany and D.~Tong, {\it {Vortices, instantons and branes}},  {\em J. High
  Energy Phys.} {\bf 07} (2003) 037,
  [\href{http://arxiv.org/abs/hep-th/0306150}{{\tt hep-th/0306150}}].

\bibitem{Auzzi:2003fs}
R.~Auzzi, S.~Bolognesi, J.~Evslin, K.~Konishi, and A.~Yung, {\it {NonAbelian
  superconductors: Vortices and confinement in N=2 SQCD}},  {\em Nucl. Phys. B}
  {\bf 673} (2003) 187--216, [\href{http://arxiv.org/abs/hep-th/0307287}{{\tt
  hep-th/0307287}}].

\bibitem{Hanany:2004ea}
A.~Hanany and D.~Tong, {\it {Vortex strings and four-dimensional gauge
  dynamics}},  {\em J. High Energy Phys.} {\bf 04} (2004) 066,
  [\href{http://arxiv.org/abs/hep-th/0403158}{{\tt hep-th/0403158}}].

\bibitem{Shifman:2004dr}
M.~Shifman and A.~Yung, {\it {NonAbelian string junctions as confined
  monopoles}},  {\em Phys. Rev. D} {\bf 70} (2004) 045004,
  [\href{http://arxiv.org/abs/hep-th/0403149}{{\tt hep-th/0403149}}].

\bibitem{Eto:2004rz}
M.~Eto, Y.~Isozumi, M.~Nitta, K.~Ohashi, and N.~Sakai, {\it {Instantons in the
  Higgs phase}},  {\em Phys. Rev. D} {\bf 72} (2005) 025011,
  [\href{http://arxiv.org/abs/hep-th/0412048}{{\tt hep-th/0412048}}].

\bibitem{Eto:2005yh}
M.~Eto, Y.~Isozumi, M.~Nitta, K.~Ohashi, and N.~Sakai, {\it {Moduli space of
  non-Abelian vortices}},  {\em Phys. Rev. Lett.} {\bf 96} (2006) 161601,
  [\href{http://arxiv.org/abs/hep-th/0511088}{{\tt hep-th/0511088}}].

\bibitem{Tong:2005un}
D.~Tong, {\it {TASI lectures on solitons: Instantons, monopoles, vortices and
  kinks}},  in {\em {Theoretical Advanced Study Institute in Elementary
  Particle Physics}: {Many Dimensions of String Theory}}, 6, 2005.
\newblock \href{http://arxiv.org/abs/hep-th/0509216}{{\tt hep-th/0509216}}.

\bibitem{Eto:2006pg}
M.~Eto, Y.~Isozumi, M.~Nitta, K.~Ohashi, and N.~Sakai, {\it {Solitons in the
  Higgs phase: The Moduli matrix approach}},  {\em J. Phys. A} {\bf 39} (2006)
  R315--R392, [\href{http://arxiv.org/abs/hep-th/0602170}{{\tt
  hep-th/0602170}}].

\bibitem{Shifman:2007ce}
M.~Shifman and A.~Yung, {\it {Supersymmetric Solitons and How They Help Us
  Understand Non-Abelian Gauge Theories}},  {\em Rev. Mod. Phys.} {\bf 79}
  (2007) 1139, [\href{http://arxiv.org/abs/hep-th/0703267}{{\tt
  hep-th/0703267}}].

\bibitem{Shifman:2009zz}
M.~Shifman and A.~Yung, {\em {Supersymmetric solitons}}.
\newblock Cambridge Monographs on Mathematical Physics. Cambridge University
  Press, 5, 2009.

\bibitem{Eto:2018hhg}
M.~Eto, M.~Kurachi, and M.~Nitta, {\it {Constraints on two Higgs doublet models
  from domain walls}},  {\em Phys. Lett. B} {\bf 785} (2018) 447--453,
  [\href{http://arxiv.org/abs/1803.04662}{{\tt arXiv:1803.04662}}].

\bibitem{Eto:2018tnk}
M.~Eto, M.~Kurachi, and M.~Nitta, {\it {Non-Abelian strings and domain walls in
  two Higgs doublet models}},  {\em J. High Energy Phys.} {\bf 08} (2018) 195,
  [\href{http://arxiv.org/abs/1805.07015}{{\tt arXiv:1805.07015}}].

\bibitem{Eto:2019hhf}
M.~Eto, Y.~Hamada, M.~Kurachi, and M.~Nitta, {\it {Topological Nambu monopole
  in two Higgs doublet models}},  {\em Phys. Lett. B} {\bf 802} (2020) 135220,
  [\href{http://arxiv.org/abs/1904.09269}{{\tt arXiv:1904.09269}}].

\bibitem{Eto:2020hjb}
M.~Eto, Y.~Hamada, M.~Kurachi, and M.~Nitta, {\it {Dynamics of Nambu monopole
  in two Higgs doublet models. Cosmological Monopole Collider}},  {\em J. High
  Energy Phys.} {\bf 07} (2020) 004,
  [\href{http://arxiv.org/abs/2003.08772}{{\tt arXiv:2003.08772}}].

\bibitem{Eto:2020opf}
M.~Eto, Y.~Hamada, and M.~Nitta, {\it {Topological structure of a Nambu
  monopole in two-Higgs-doublet models: Fiber bundle, Dirac\textquoteright{}s
  quantization, and a dyon}},  {\em Phys. Rev. D} {\bf 102} (2020), no.~10
  105018, [\href{http://arxiv.org/abs/2007.15587}{{\tt arXiv:2007.15587}}].

\bibitem{Eto:2021dca}
M.~Eto, Y.~Hamada, and M.~Nitta, {\it {Stable Z-strings with topological
  polarization in two Higgs doublet model}},  {\em J. High Energy Phys.} {\bf
  02} (2022) 099, [\href{http://arxiv.org/abs/2111.13345}{{\tt
  arXiv:2111.13345}}].

\bibitem{Forbes:2001gj}
M.~M. Forbes and A.~R. Zhitnitsky, {\it {Global strings in high density QCD}},
  {\em Phys. Rev. D} {\bf 65} (2002) 085009,
  [\href{http://arxiv.org/abs/hep-ph/0109173}{{\tt hep-ph/0109173}}].

\bibitem{Iida:2002ev}
K.~Iida and G.~Baym, {\it {Superfluid phases of quark matter. 3. Supercurrents
  and vortices}},  {\em Phys. Rev. D} {\bf 66} (2002) 014015,
  [\href{http://arxiv.org/abs/hep-ph/0204124}{{\tt hep-ph/0204124}}].

\bibitem{Cipriani:2012hr}
M.~Cipriani, W.~Vinci, and M.~Nitta, {\it {Colorful boojums at the interface of
  a color superconductor}},  {\em Phys. Rev. D} {\bf 86} (2012) 121704,
  [\href{http://arxiv.org/abs/1208.5704}{{\tt arXiv:1208.5704}}].

\bibitem{Alford:2016dco}
M.~G. Alford, S.~Mallavarapu, T.~Vachaspati, and A.~Windisch, {\it {Stability
  of superfluid vortices in dense quark matter}},  {\em Phys. Rev. C} {\bf 93}
  (2016), no.~4 045801, [\href{http://arxiv.org/abs/1601.04656}{{\tt
  arXiv:1601.04656}}].

\bibitem{Yasui:2010yw}
S.~Yasui, K.~Itakura, and M.~Nitta, {\it {Fermion structure of non-Abelian
  vortices in high density QCD}},  {\em Phys. Rev. D} {\bf 81} (2010) 105003,
  [\href{http://arxiv.org/abs/1001.3730}{{\tt arXiv:1001.3730}}].

\bibitem{Fujiwara:2011za}
T.~Fujiwara, T.~Fukui, M.~Nitta, and S.~Yasui, {\it {Index theorem and Majorana
  zero modes along a non-Abelian vortex in a color superconductor}},  {\em
  Phys. Rev. D} {\bf 84} (2011) 076002,
  [\href{http://arxiv.org/abs/1105.2115}{{\tt arXiv:1105.2115}}].

\bibitem{Chatterjee:2016ykq}
C.~Chatterjee, M.~Cipriani, and M.~Nitta, {\it {Coupling between Majorana
  fermions and Nambu-Goldstone bosons inside a non-Abelian vortex in dense
  QCD}},  {\em Phys. Rev. D} {\bf 93} (2016), no.~6 065046,
  [\href{http://arxiv.org/abs/1602.01677}{{\tt arXiv:1602.01677}}].

\bibitem{Kobayashi:2013axa}
M.~Kobayashi, E.~Nakano, and M.~Nitta, {\it {Color Magnetism in Non-Abelian
  Vortex Matter}},  {\em J. High Energy Phys.} {\bf 06} (2014) 130,
  [\href{http://arxiv.org/abs/1311.2399}{{\tt arXiv:1311.2399}}].

\bibitem{Hirono:2012ki}
Y.~Hirono and M.~Nitta, {\it {Anisotropic optical response of dense quark
  matter under rotation: Compact stars as cosmic polarizers}},  {\em Phys. Rev.
  Lett.} {\bf 109} (2012) 062501, [\href{http://arxiv.org/abs/1203.5059}{{\tt
  arXiv:1203.5059}}].

\bibitem{Alford:2018mqj}
M.~G. Alford, G.~Baym, K.~Fukushima, T.~Hatsuda, and M.~Tachibana, {\it
  {Continuity of vortices from the hadronic to the color-flavor locked phase in
  dense matter}},  {\em Phys. Rev. D} {\bf 99} (2019), no.~3 036004,
  [\href{http://arxiv.org/abs/1803.05115}{{\tt arXiv:1803.05115}}].

\bibitem{Chatterjee:2018nxe}
C.~Chatterjee, M.~Nitta, and S.~Yasui, {\it {Quark-hadron continuity under
  rotation: Vortex continuity or boojum?}},  {\em Phys. Rev. D} {\bf 99}
  (2019), no.~3 034001, [\href{http://arxiv.org/abs/1806.09291}{{\tt
  arXiv:1806.09291}}].

\bibitem{Chatterjee:2019tbz}
C.~Chatterjee, M.~Nitta, and S.~Yasui, {\it {Quark-Hadron Crossover with
  Vortices}},  {\em JPS Conf. Proc.} {\bf 26} (2019) 024030,
  [\href{http://arxiv.org/abs/1902.00156}{{\tt arXiv:1902.00156}}].

\bibitem{Cherman:2018jir}
A.~Cherman, S.~Sen, and L.~G. Yaffe, {\it {Anyonic particle-vortex statistics
  and the nature of dense quark matter}},  {\em Phys. Rev. D} {\bf 100} (2019),
  no.~3 034015, [\href{http://arxiv.org/abs/1808.04827}{{\tt
  arXiv:1808.04827}}].

\bibitem{Hirono:2018fjr}
Y.~Hirono and Y.~Tanizaki, {\it {Quark-Hadron Continuity beyond the
  Ginzburg-Landau Paradigm}},  {\em Phys. Rev. Lett.} {\bf 122} (2019), no.~21
  212001, [\href{http://arxiv.org/abs/1811.10608}{{\tt arXiv:1811.10608}}].

\bibitem{Hirono:2019oup}
Y.~Hirono and Y.~Tanizaki, {\it {Effective gauge theories of superfluidity with
  topological order}},  {\em J. High Energy Phys.} {\bf 07} (2019) 062,
  [\href{http://arxiv.org/abs/1904.08570}{{\tt arXiv:1904.08570}}].

\bibitem{Cherman:2020hbe}
A.~Cherman, T.~Jacobson, S.~Sen, and L.~G. Yaffe, {\it {Higgs-confinement phase
  transitions with fundamental representation matter}},  {\em Phys. Rev. D}
  {\bf 102} (2020), no.~10 105021, [\href{http://arxiv.org/abs/2007.08539}{{\tt
  arXiv:2007.08539}}].

\bibitem{Hayashi:2023sas}
Y.~Hayashi, {\it {Higgs-Confinement Continuity and Matching of Aharonov-Bohm
  Phases}},  {\em Phys. Rev. Lett.} {\bf 132} (2024), no.~22 221901,
  [\href{http://arxiv.org/abs/2303.02129}{{\tt arXiv:2303.02129}}].

\bibitem{Cherman:2024exo}
A.~Cherman, T.~Jacobson, S.~Sen, and L.~G. Yaffe, {\it {Line operators, vortex
  statistics, and Higgs versus confinement dynamics}},  {\em J. High Energy
  Phys.} {\bf 06} (2024) 200, [\href{http://arxiv.org/abs/2401.17489}{{\tt
  arXiv:2401.17489}}].

\bibitem{Hayata:2024nrl}
T.~Hayata, Y.~Hidaka, and D.~Kondo, {\it {Phase transition on superfluid
  vortices in Higgs-Confinement crossover}},
  \href{http://arxiv.org/abs/2411.03676}{{\tt arXiv:2411.03676}}.

\bibitem{Babaev:2001hv}
E.~Babaev, {\it {Vortices carrying an arbitrary fraction of magnetic flux
  quantum in two gap superconductors}},  {\em Phys. Rev. Lett.} {\bf 89} (2002)
  067001, [\href{http://arxiv.org/abs/cond-mat/0111192}{{\tt
  cond-mat/0111192}}].

\bibitem{doi:10.1143/JPSJ.70.2844}
Y.~Tanaka, {\it Phase instability in multi-band superconductors},  {\em Journal
  of the Physical Society of Japan} {\bf 70} (2001), no.~10 2844--2847.

\bibitem{PhysRevLett.88.017002}
Y.~Tanaka, {\it Soliton in two-band superconductor},  {\em Phys. Rev. Lett.}
  {\bf 88} (Dec, 2001) 017002.

\bibitem{Goryo_2007}
J.~Goryo, S.~Soma, and H.~Matsukawa, {\it Deconfinement of vortices with
  continuously variable fractions of the unit flux quanta in two-gap
  superconductors},  {\em Europhysics Letters ({EPL})} {\bf 80} (sep, 2007)
  17002.

\bibitem{PhysRevLett.80.5184}
D.~F. Agterberg, {\it Vortex lattice structures of
  ${\mathrm{sr}}_{2}{\mathrm{ruo}}_{4}$},  {\em Phys. Rev. Lett.} {\bf 80}
  (Jun, 1998) 5184--5187.

\bibitem{PhysRevB.86.060514}
J.~Garaud and E.~Babaev, {\it Skyrmionic state and stable half-quantum vortices
  in chiral $p$-wave superconductors},  {\em Phys. Rev. B} {\bf 86} (Aug, 2012)
  060514.

\bibitem{Son:2001td}
D.~T. Son and M.~A. Stephanov, {\it {Domain walls in two-component
  Bose-Einstein condensates}},  {\em Phys. Rev. A} {\bf 65} (2002) 063621,
  [\href{http://arxiv.org/abs/cond-mat/0103451}{{\tt cond-mat/0103451}}].

\bibitem{Kasamatsu:2004tvg}
K.~Kasamatsu, M.~Tsubota, and M.~Ueda, {\it {Vortex molecules in coherently
  coupled two-component Bose-Einstein condensates}},  {\em Phys. Rev. Lett.}
  {\bf 93} (2004), no.~25 250406,
  [\href{http://arxiv.org/abs/cond-mat/0406150}{{\tt cond-mat/0406150}}].

\bibitem{Cipriani:2013nya}
M.~Cipriani and M.~Nitta, {\it {Crossover between integer and fractional vortex
  lattices in coherently coupled two-component Bose-Einstein condensates}},
  {\em Phys. Rev. Lett.} {\bf 111} (2013) 170401,
  [\href{http://arxiv.org/abs/1303.2592}{{\tt arXiv:1303.2592}}].

\bibitem{Tylutki:2016mgy}
M.~Tylutki, L.~P. Pitaevskii, A.~Recati, and S.~Stringari, {\it {Confinement
  and precession of vortex pairs in coherently coupled Bose-Einstein
  condensates}},  {\em Phys. Rev. A} {\bf 93} (2016), no.~4 043623,
  [\href{http://arxiv.org/abs/1601.03695}{{\tt arXiv:1601.03695}}].

\bibitem{Eto:2017rfr}
M.~Eto and M.~Nitta, {\it {Confinement of half-quantized vortices in coherently
  coupled Bose-Einstein condensates: Simulating quark confinement in a QCD-like
  theory}},  {\em Phys. Rev. A} {\bf 97} (2018), no.~2 023613,
  [\href{http://arxiv.org/abs/1702.04892}{{\tt arXiv:1702.04892}}].

\bibitem{Eto:2019uhe}
M.~Eto, K.~Ikeno, and M.~Nitta, {\it {Collision dynamics and reactions of
  fractional vortex molecules in coherently coupled Bose-Einstein
  condensates}},  {\em Phys. Rev. Res.} {\bf 2} (2020), no.~3 033373,
  [\href{http://arxiv.org/abs/1912.09014}{{\tt arXiv:1912.09014}}].

\bibitem{Kobayashi:2018ezm}
M.~Kobayashi, M.~Eto, and M.~Nitta, {\it {Berezinskii-Kosterlitz-Thouless
  Transition of Two-Component Bose Mixtures with Intercomponent Josephson
  Coupling}},  {\em Phys. Rev. Lett.} {\bf 123} (2019), no.~7 075303,
  [\href{http://arxiv.org/abs/1802.08763}{{\tt arXiv:1802.08763}}].

\bibitem{Chatterjee:2019jez}
C.~Chatterjee, S.~B. Gudnason, and M.~Nitta, {\it {Chemical bonds of two vortex
  species with a generalized Josephson term and arbitrary charges}},  {\em J.
  High Energy Phys.} {\bf 04} (2020) 109,
  [\href{http://arxiv.org/abs/1912.02685}{{\tt arXiv:1912.02685}}].

\bibitem{Eto:2012rc}
M.~Eto and M.~Nitta, {\it {Vortex trimer in three-component Bose-Einstein
  condensates}},  {\em Phys. Rev. A} {\bf 85} (2012) 053645,
  [\href{http://arxiv.org/abs/1201.0343}{{\tt arXiv:1201.0343}}].

\bibitem{Eto:2013spa}
M.~Eto and M.~Nitta, {\it {Vortex graphs as N-omers and CP(N-1) Skyrmions in
  N-component Bose-Einstein condensates}},  {\em EPL} {\bf 103} (2013), no.~6
  60006, [\href{http://arxiv.org/abs/1303.6048}{{\tt arXiv:1303.6048}}].

\bibitem{Eto:2021nle}
M.~Eto and M.~Nitta, {\it {Chiral non-Abelian vortices and their confinement in
  three flavor dense QCD}},  {\em Phys. Rev. D} {\bf 104} (2021), no.~9 094052,
  [\href{http://arxiv.org/abs/2103.13011}{{\tt arXiv:2103.13011}}].

\bibitem{Eto:2013bxa}
M.~Eto, Y.~Hirono, and M.~Nitta, {\it {Domain Walls and Vortices in Chiral
  Symmetry Breaking}},  {\em PTEP} {\bf 2014} (2014), no.~3 033B01,
  [\href{http://arxiv.org/abs/1309.4559}{{\tt arXiv:1309.4559}}].

\bibitem{Iida:2000ha}
K.~Iida and G.~Baym, {\it {The Superfluid phases of quark matter:
  Ginzburg-Landau theory and color neutrality}},  {\em Phys. Rev. D} {\bf 63}
  (2001) 074018, [\href{http://arxiv.org/abs/hep-ph/0011229}{{\tt
  hep-ph/0011229}}]. [Erratum: Phys.Rev.D 66, 059903 (2002)].

\bibitem{Iida:2001pg}
K.~Iida and G.~Baym, {\it {Superfluid phases of quark matter. 2: phenomenology
  and sum rules}},  {\em Phys. Rev. D} {\bf 65} (2002) 014022,
  [\href{http://arxiv.org/abs/hep-ph/0108149}{{\tt hep-ph/0108149}}].

\bibitem{Giannakis:2001wz}
I.~Giannakis and H.-c. Ren, {\it {The Ginzburg-Landau free energy functional of
  color superconductivity at weak coupling}},  {\em Phys. Rev. D} {\bf 65}
  (2002) 054017, [\href{http://arxiv.org/abs/hep-ph/0108256}{{\tt
  hep-ph/0108256}}].

\bibitem{Nitta:2014rxa}
M.~Nitta, {\it {Non-Abelian Sine-Gordon Solitons}},  {\em Nucl. Phys. B} {\bf
  895} (2015) 288--302, [\href{http://arxiv.org/abs/1412.8276}{{\tt
  arXiv:1412.8276}}].

\bibitem{Eto:2015uqa}
M.~Eto and M.~Nitta, {\it {Non-Abelian Sine-Gordon Solitons: Correspondence
  between $SU(N)$ Skyrmions and ${\mathbb C}P^{N-1}$ Lumps}},  {\em Phys. Rev.
  D} {\bf 91} (2015), no.~8 085044,
  [\href{http://arxiv.org/abs/1501.07038}{{\tt arXiv:1501.07038}}].

\bibitem{Eto:2021gyy}
M.~Eto, K.~Nishimura, and M.~Nitta, {\it {Phases of rotating baryonic matter:
  non-Abelian chiral soliton lattices, antiferro-isospin chains, and
  ferri/ferromagnetic magnetization}},  {\em J. High Energy Phys.} {\bf 08}
  (2022) 305, [\href{http://arxiv.org/abs/2112.01381}{{\tt arXiv:2112.01381}}].

\bibitem{Nitta:2015mma}
M.~Nitta, {\it {Josephson junction of non-Abelian superconductors and
  non-Abelian Josephson vortices}},  {\em Nucl. Phys. B} {\bf 899} (2015)
  78--90, [\href{http://arxiv.org/abs/1502.02525}{{\tt arXiv:1502.02525}}].

\bibitem{Nitta:2015mxa}
M.~Nitta, {\it {Josephson instantons and Josephson monopoles in a non-Abelian
  Josephson junction}},  {\em Phys. Rev. D} {\bf 92} (2015), no.~4 045010,
  [\href{http://arxiv.org/abs/1503.02060}{{\tt arXiv:1503.02060}}].

\bibitem{Nitta:2022ahj}
M.~Nitta, {\it {Relations among topological solitons}},  {\em Phys. Rev. D}
  {\bf 105} (2022), no.~10 105006, [\href{http://arxiv.org/abs/2202.03929}{{\tt
  arXiv:2202.03929}}].

\bibitem{Eto:2023tuu}
M.~Eto, K.~Nishimura, and M.~Nitta, {\it {Domain-wall Skyrmion phase in a
  rapidly rotating QCD matter}},  {\em J. High Energy Phys.} {\bf 03} (2024)
  019, [\href{http://arxiv.org/abs/2310.17511}{{\tt arXiv:2310.17511}}].

\bibitem{Amari:2024jxx}
Y.~Amari and M.~Nitta, {\it {Skyrmion crystal phase on a magnetic domain wall
  in chiral magnets}},  \href{http://arxiv.org/abs/2409.07943}{{\tt
  arXiv:2409.07943}}.

\bibitem{Speight:1996px}
J.~M. Speight, {\it {Static intervortex forces}},  {\em Phys. Rev. D} {\bf 55}
  (1997) 3830--3835, [\href{http://arxiv.org/abs/hep-th/9603155}{{\tt
  hep-th/9603155}}].

\bibitem{Vinci:2012mc}
W.~Vinci, M.~Cipriani, and M.~Nitta, {\it {Spontaneous Magnetization through
  Non-Abelian Vortex Formation in Rotating Dense Quark Matter}},  {\em Phys.
  Rev. D} {\bf 86} (2012) 085018, [\href{http://arxiv.org/abs/1206.3535}{{\tt
  arXiv:1206.3535}}].

\bibitem{Chatterjee:2015lbf}
C.~Chatterjee and M.~Nitta, {\it {Aharonov-Bohm Phase in High Density Quark
  Matter}},  {\em Phys. Rev. D} {\bf 93} (2016), no.~6 065050,
  [\href{http://arxiv.org/abs/1512.06603}{{\tt arXiv:1512.06603}}].

\bibitem{Fujimoto:2020dsa}
Y.~Fujimoto and M.~Nitta, {\it {Non-Abelian Alice strings in two-flavor dense
  QCD}},  {\em Phys. Rev. D} {\bf 103} (2021) 054002,
  [\href{http://arxiv.org/abs/2011.09947}{{\tt arXiv:2011.09947}}].

\bibitem{Fujimoto:2021bes}
Y.~Fujimoto and M.~Nitta, {\it {Alice meets Boojums in neutron stars: vortices
  penetrating two-flavor quark-hadron continuity}},  {\em Phys. Rev. D} {\bf
  103} (2021), no.~11 114003, [\href{http://arxiv.org/abs/2102.12928}{{\tt
  arXiv:2102.12928}}].

\bibitem{Fujimoto:2021wsr}
Y.~Fujimoto and M.~Nitta, {\it {Topological confinement of vortices in
  two-flavor dense QCD}},  {\em J. High Energy Phys.} {\bf 09} (2021) 192,
  [\href{http://arxiv.org/abs/2103.15185}{{\tt arXiv:2103.15185}}].

\bibitem{Yasui:2010yh}
S.~Yasui, K.~Itakura, and M.~Nitta, {\it {Majorana meets Coxeter: Non-Abelian
  Majorana Fermions and Non-Abelian Statistics}},  {\em Phys. Rev. B} {\bf 83}
  (2011) 134518, [\href{http://arxiv.org/abs/1010.3331}{{\tt
  arXiv:1010.3331}}].

\bibitem{Hirono:2012ad}
Y.~Hirono, S.~Yasui, K.~Itakura, and M.~Nitta, {\it {Non-Abelian statistics of
  vortices with multiple Majorana fermions}},  {\em Phys. Rev. B} {\bf 86}
  (2012) 014508, [\href{http://arxiv.org/abs/1203.0173}{{\tt
  arXiv:1203.0173}}].

\bibitem{Masaki:2023rtn}
Y.~Masaki, T.~Mizushima, and M.~Nitta, {\it {Non-Abelian Anyons and Non-Abelian
  Vortices in Topological Superconductors}},  {\em Encyclopedia of Condensed
  Matter Physics (Second Edition)} {\bf 2} (1, 2024) 755--794,
  [\href{http://arxiv.org/abs/2301.11614}{{\tt arXiv:2301.11614}}].

\end{thebibliography}\endgroup

\end{document}